\documentclass[11pt,a4paper]{article}

\usepackage[a4paper,margin=1in]{geometry}
\usepackage{graphicx}
\usepackage{amsmath}
\usepackage{amssymb}
\usepackage{booktabs}
\usepackage{color}
\usepackage{rotating}
\usepackage{caption}
\usepackage{ulem}
\usepackage{soul}
\usepackage[numbers,sort&compress]{natbib}
\usepackage[hidelinks]{hyperref}
\usepackage{url}
\hypersetup{%
  hidelinks,
  pdfborder={0 0 0},
  pdftitle={Deep Learning with Magnetic Parameter Constraints for Short-Term Prediction of Solar Active Region Vector Magnetic Fields},
  pdfauthor={Yuqing Zhou, Hui Liu, Zhenyu Jin, Yuyang Li, Sizhong Zou, Jiaben Lin, Mingfu Shao, and Zhuoheng Huang},
  pdfkeywords={Solar magnetic fields; Neural networks; Active region magnetic fields; Space weather; Deep learning}
}

\graphicspath{{./}{pic/}{pic3/}{pic4/}{pic_appendix/}}

\title{Deep Learning with Magnetic Parameter Constraints for Short-Term Prediction of Solar Active Region Vector Magnetic Fields}

\author{%
Yuqing Zhou$^{1,2}$,
Hui Liu$^{1,2}$\thanks{Corresponding author: \texttt{liuhui@ynao.ac.cn}},
Zhenyu Jin$^{1,2}$,
Yuyang Li$^{2,3}$,\\
Sizhong Zou$^{1,2}$,
Jiaben Lin$^{2,3}$,
Mingfu Shao$^{2,3}$,
and Zhuoheng Huang$^{1,2}$\\[0.8em]
\small $^1$Yunnan Observatories, Chinese Academy of Sciences, Kunming 650216, People's Republic of China\\
\small $^2$University of Chinese Academy of Sciences, Beijing 100049, People's Republic of China\\
\small $^3$National Astronomical Observatories, Chinese Academy of Sciences, Beijing 100101, People's Republic of China
}

\date{}

\begin{document}

\maketitle

\begin{abstract}
Forecasting the dynamic evolution of solar magnetic fields is a critical technique for enabling space weather warnings. Addressing the limitations of existing methods in predicting all vector magnetic field components and in maintaining consistency with solar surface magnetic-field-related quantities, this study proposes a deep learning prediction method that integrates dynamic masks of active regions with multiple magnetic parameter constraints. By constructing a three-channel representation of vector magnetic fields, applying dynamic masks to enhance attention to strong-field regions, and incorporating multi-parameter magnetic parameter constraints, we developed an end-to-end short-term (12-hour) predictive model of solar vector magnetic field evolution. Using SDO/SHARP vector magnetogram data, the model predicts and analyses field evolution across all components. Quantitative evaluations demonstrate that our approach achieves horizon-averaged structural similarity index measure (SSIM) of 0.912 (per-hour range: 0.909--0.916) and correlation coefficient (CC) of 0.998 for the radial component $B_r$ (root-mean-square error (RMSE) 13.0--21.0~G); the horizontal components achieve $B_\phi$ SSIM 0.760--0.800 (CC 0.910--0.945, RMSE 38.5--50.0~G) and $B_\theta$ SSIM 0.728--0.750 (CC 0.895--0.920, RMSE 38.5--49.0~G). The model maintains unsigned magnetic flux prediction errors at 7.82\% (95\% confidence interval (CI): $\pm$0.11\%). These results demonstrate strong image-domain performance together with consistency under the magnetic-parameter diagnostics used here, suggesting initial potential for supporting future space weather forecasting efforts.
\end{abstract}


\section{Introduction}
The ability to forecast solar magnetic field evolution is crucial for improving the accuracy and timeliness of space weather warnings. In recent years, significant progress has been made in both physics-based modeling and data-driven approaches:

Physics-based modeling has made substantial progress: local reconstruction of active region magnetic fields has revealed precursors of sudden X-10 class flares through strong shear flows along polarity inversion lines (PILs) \citep{yang2004}.  
Surface flux transport (SFT) models focus primarily on the radial component \citep{jiang2014b}, often neglecting horizontal components critical for magnetic reconnection and shear. Jiang et al. \citep{jiang2014} used a Monte Carlo inclination divergence method to simulate large-scale field evolution, while oversimplification of horizontal components in transport models can lead to biases in short-term polar field variations \citep{jeong2025b}.

Data-driven approaches have also advanced rapidly. Covas et al.\ \citep{covas2019} applied spatiotemporal neural network embeddings to forecast the full sunspot diagram (latitude--time), not magnetic fields. Bai et al. \citep{bai2021} developed deep learning models targeting single components.

Photospheric vector magnetic fields provide critical information for characterizing the free magnetic energy and current systems that drive eruptive solar events \citep{Toriumi2019, bobra2015}, making their accurate prediction essential for space weather applications.

Despite these advances, existing methods face key limitations in meeting the high standards of space weather warnings that require complete vector representation and consistency with solar surface magnetic-field-related quantities:

\begin{enumerate}
    \item \textbf{Incomplete vector field prediction.} Upton \& Hathaway's model \citep{upton2014} handled only the radial component while ignoring horizontal fields. SFT models \citep{jeong2025b} simplify horizontal components, causing biases in polar field evolution. Data-driven models such as Bai et al.\ \citep{bai2021} focus on single-component radial forecasting; Covas et al.\ \citep{covas2019} model sunspot spatiotemporal diagrams rather than magnetic fields. Jeong et al.\ \citep{jeong2025a} introduced a Pix2PixCC-based model predicting full-disk radial-field evolution at flexible time steps up to 27 days, without horizontal outputs, limiting calculation of horizontal-field-dependent quantities such as current helicity. Ramunno et al. \citep{ramunno2024} applied diffusion probabilistic models to predict full-disk line-of-sight magnetograms but still lack complete vector outputs needed to characterize three-dimensional topology.
    \item \textbf{Limited magnetic-parameter consistency.} Data-driven models such as Bai et al.\ \citep{bai2021} often omit sufficient physical constraints, risking outputs that violate flux conservation or Lorentz force balance; Covas et al.\ \citep{covas2020} similarly apply transfer learning to sunspot and longitude-averaged radial field data without explicit magnetic-parameter constraints. Jeong et al.\ \citep{jeong2025a} introduced a CC-based consistency loss but excluded constraints such as Lorentz force balance and helicity conservation. Ramunno et al. \citep{ramunno2024} included checks such as flux and AR area without cross-scale consistency. Physics-based models, though consistent with theory, often restrict themselves to single-parameter constraints \citep{fan2015} and lack multi-parameter integration.
\end{enumerate}

Current approaches are therefore insufficient for reliable space weather forecasting that requires both high prediction accuracy and consistency with solar surface magnetic-field-related quantities. This limitation increases the risk of mischaracterizing the severity of coronal mass ejections \citep{kilpua2019}.

To address these challenges, we propose a magnetic-parameter-constrained deep learning architecture that integrates dynamic active region masks with multi-parameter magnetic parameter constraints. The approach enhances feature extraction from critical structures such as polarity inversion lines and sunspots via dynamic masks, while incorporating joint constraints on solar surface magnetic-field-related quantities including total magnetic pressure and unsigned magnetic flux. Using SDO/SHARP vector magnetogram data, the model learns spatiotemporal connections in field evolution: given a 12-hour input sequence, it forecasts the subsequent 12 hours of evolution across all components while maintaining both topological fidelity and parameter consistency. Quantitative evaluations show that the method achieves structural similarity of 0.909--0.916 (horizon average $\approx$0.912) for the radial component (CC 0.997--0.998, RMSE 13.0--21.0~G), with unsigned flux errors of 7.82\% (95\% CI: ±0.11\%). These results suggest that the proposed approach represents an initial exploration toward future operational space weather forecasting.

\section{Deep Learning Method with Dynamic Mask Enhancement and Magnetic Parameter Constraints}
We propose a prediction method that integrates dynamic active region masks with magnetic parameter constraints into an end-to-end deep learning architecture. This design enhances the model's ability to learn individual vector magnetic field components while maintaining vector magnetic parameter consistency.

\subsection{Attention-weighted Strategy for Active Region Cores}
To comprehensively represent active region vector fields, the input data are constructed in Carrington heliographic projection as three channels corresponding to the radial ($B_r$), phi ($B_\phi$, also denoted $B_p$), and theta ($B_\theta$, also denoted $B_t$) components, plus one mask channel, yielding a four-channel input tensor.

To capture critical structures, we adopt a two-stage mask construction mechanism guided by the DSARD (Deep-learning-based Solar Active Region Detection; \citet{chen2025}) active region detection method. First, static masks are generated from the input 12-frame sequence based on the $B_r$ channel: thresholding eliminates quiet-Sun noise, while DBSCAN-like (Density-Based Spatial Clustering of Applications with Noise) clustering builds coherent mask regions. This combines threshold efficiency with morphological adaptability, preventing fragmentation and better delineating critical features such as shear bands near polarity inversion lines.  
During training, masks are dynamically predicted alongside magnetic field evolution using temporal gradients, producing $\mathbf{M}_t$ where core strong-field regions are 1 and background regions are 0. This enforces stronger attention on magnetically active regions, particularly near polarity inversion lines where magnetic shear and gradients are steepest.

\subsection{Incorporation and Implementation of Magnetic Parameter Constraints}
To provide magnetically meaningful inductive bias, we condition the network on five differentiable magnetic-parameter-based diagnostic maps computed directly from the vector magnetogram at each time step. These maps are injected as auxiliary channels to guide feature extraction, while the learning objective remains defined on the predicted magnetograms themselves.

\paragraph{Rationale.} While convolutional neural networks can in principle learn to compute spatial derivatives and other low-level operations through training, explicitly providing these magnetic-parameter-based features offers several advantages: it reduces the learning burden on early network layers, allowing them to focus on higher-level spatiotemporal patterns; it ensures numerical stability and consistency by using fixed, differentiable implementations; and it provides an interpretable conditioning pathway that can be systematically ablated. These auxiliary channels serve as structured inductive biases that guide the learning process toward magnetically consistent solutions.

\paragraph{Magnetic-parameter-based diagnostic maps.} The five auxiliary feature maps are:

\begin{itemize}
    \item \textbf{Lorentz-force proxy (magnetic pressure):}
    \begin{equation}
    P_{\mathrm{mag}} = B_r^2 + B_p^2 + B_t^2,
    \end{equation}
    which serves as a computationally stable proxy for magnetic energy density. The full Lorentz force is $\mathbf{F}_L = \mathbf{J} \times \mathbf{B}$. Since computing the full three-dimensional current density requires additional derivatives and assumptions about vertical structure, we use magnetic pressure as a scalar proxy that highlights regions where magnetic forces are dynamically important \citep{fisher2012}.
    \item \textbf{Unsigned flux:}
    \begin{equation}
    |\Phi| = |B_r|,
    \end{equation}
    which directly measures the unsigned radial flux density \citep{leka2003}.
    \item \textbf{Shear-angle proxy:}
    \begin{equation}
    \alpha = \arccos\left(\frac{|B_r|}{|B_r|+\sqrt{B_p^2+B_t^2}+\epsilon}\right),\quad \epsilon=10^{-6},
    \end{equation}
    which increases when transverse fields are relatively strong, highlighting non-potential structure.
    \item \textbf{Radial gradient magnitude:}
    \begin{equation}
    |\nabla B_r| = \sqrt{\left(\frac{\partial B_r}{\partial x}\right)^2 + \left(\frac{\partial B_r}{\partial y}\right)^2},
    \end{equation}
    computed using Sobel operators to emphasize steep-gradient boundaries \citep{cui2007}.
    \item \textbf{Horizontal gradient magnitude:}
    \begin{equation}
    |\nabla B_h| = \sqrt{|\nabla B_p|^2+|\nabla B_t|^2},
    \end{equation}
    where $|\nabla B_p|$ and $|\nabla B_t|$ are computed by Sobel operators.
\end{itemize}

These diagnostic maps are combined into a five-channel tensor:
\begin{equation}
\mathbf{P}_t = [P_{\mathrm{mag}},\ |\Phi|,\ \alpha,\ |\nabla B_r|,\ |\nabla B_h|]
\end{equation}
which is concatenated to the model input at each time step as auxiliary channels.

\paragraph{Implementation details.} The magnetic-parameter-based features are computed using a dedicated \texttt{RealtimePhysicsLayer} module implemented in PyTorch. This module was designed and implemented by the authors specifically for this study; it is not derived from any external library. The source code is publicly available in our repository at \url{https://github.com/Qingcaiyurouzhou/magnetic_predict}.

Internally, the \texttt{RealtimePhysicsLayer} accepts a three-channel tensor $(B_r, B_p, B_t)$ of shape $(B, 3, H, W)$ and returns a five-channel output $(B, 5, H, W)$. The module pre-registers two fixed $3\times3$ Sobel kernels ($\mathbf{S}_x$ and $\mathbf{S}_y = \mathbf{S}_x^{\mathsf{T}}$) as non-trainable buffers. For each input frame, it computes: (i)~magnetic pressure $P_{\mathrm{mag}} = B_r^2 + B_p^2 + B_t^2$; (ii)~unsigned flux $|\Phi| = |B_r|$; (iii)~shear angle via $\alpha = \arccos\!\bigl(|B_r|/(|B_r| + \sqrt{B_p^2 + B_t^2} + \epsilon)\bigr)$ with the argument clamped to $(-1+10^{-4},\,1-10^{-4})$ for numerical stability; (iv)~radial gradient magnitude $|\nabla B_r| = \sqrt{(\mathbf{S}_x * B_r)^2 + (\mathbf{S}_y * B_r)^2}\,/\,8$ using depthwise convolution with zero-padding; and (v)~horizontal gradient magnitude $|\nabla B_h| = \sqrt{(\mathbf{S}_x * B_p)^2 + (\mathbf{S}_y * B_p)^2 + (\mathbf{S}_x * B_t)^2 + (\mathbf{S}_y * B_t)^2}\,/\,8$. All operations are fully differentiable, require no learnable parameters, and run on GPU with negligible overhead (less than 2\% of total forward-pass time). The auxiliary feature maps are computed online at each time step from the current magnetic field state, ensuring consistency throughout the autoregressive rollout.

\subsection{Model Architecture and Training Mechanism}
\begin{figure*}[htbp]
\centering
\includegraphics[width=0.65\textwidth]{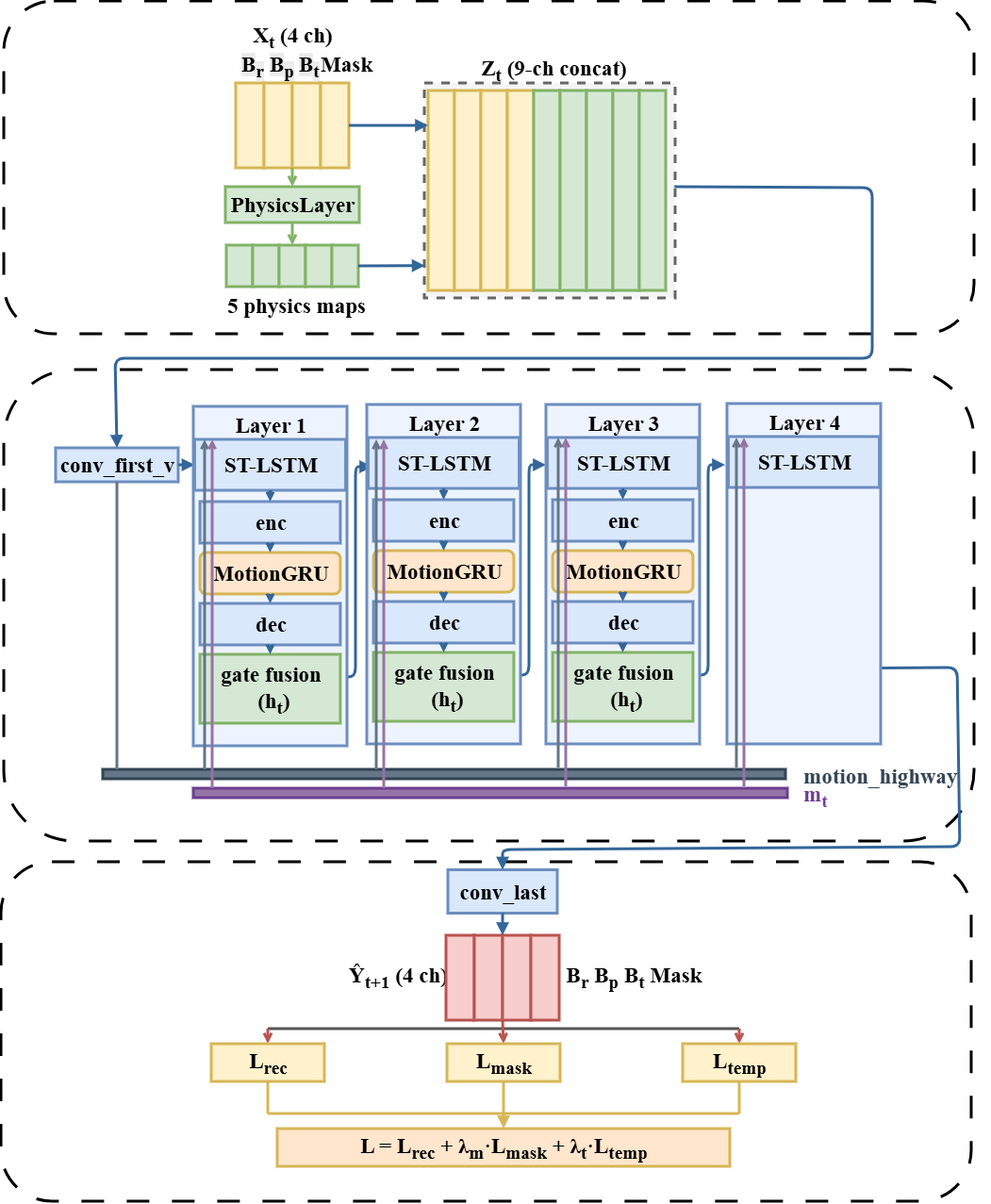}
\caption{Schematic overview of the proposed model architecture for a single time step. \textbf{Top:} The four-channel input magnetogram ($B_r$, $B_p$, $B_t$, mask) is passed to the \texttt{RealtimePhysicsLayer}, which computes five magnetic-parameter-based diagnostic maps (magnetic pressure, unsigned flux, shear-angle proxy, $|\nabla B_r|$, $|\nabla B_h|$). The nine-channel concatenation $Z_t$ is projected by \texttt{conv\_first\_v} to initialise the \texttt{motion\_highway} shared state. \textbf{Middle:} Four stacked SpatioTemporalLSTM (ST-LSTM) layers process the spatiotemporal features. In Layers~1--3, each ST-LSTM is followed by an encoder (\texttt{enc}), a MotionGRU module for local motion estimation, a decoder (\texttt{dec}), and a cross-scale gate-fusion layer that blends motion features with the hidden state $h_t^{(l)}$. Layer~4 contains only the ST-LSTM cell. The \texttt{motion\_highway} and spatial memory $m_t$ buses (shown below the layers) carry shared recurrent state sequentially across all four layers. \textbf{Bottom:} The final hidden state is projected by \texttt{conv\_last} to produce the four-channel predicted frame $\hat{Y}_{t+1}$. During autoregressive rollout (dashed feedback arrow), $\hat{Y}_{t+1}$ serves as the next input $X_{t+1}$, and the diagnostic maps are recomputed accordingly. The parallel loss branches $\mathcal{L}_{\mathrm{rec}}$, $\mathcal{L}_{\mathrm{mask}}$, and $\mathcal{L}_{\mathrm{temp}}$ are computed from $\hat{Y}_{t+1}$ and combined into the total objective $\mathcal{L}$.}
\label{fig:model_arch}
\end{figure*}

The model adapts the MotionRNN/PredRNN family \citep{wu2021} for solar applications. Each time step is represented by four channels (three magnetic components plus the active-region mask), and the auxiliary magnetic-parameter tensor $\mathbf{P}_t$ is computed online from $(B_r,B_p,B_t)$ and concatenated as additional channels. Figure~\ref{fig:model_arch} provides a schematic overview of the per-time-step forward pass, illustrating the data flow from input magnetograms through the spatiotemporal prediction network to the output predicted fields.

\paragraph{Network architecture.} The backbone consists of stacked spatiotemporal LSTM cells organized in an encoder--motion--decoder architecture with cross-scale fusion:

\begin{itemize}
    \item \textbf{Input processing:} The four-channel magnetogram is patchified (default patch size $1\times1$, yielding 4 channels per patch). The five magnetic-parameter-based diagnostic maps are computed from the magnetic components and concatenated, yielding 9 input channels per time step.
    
    \item \textbf{Spatiotemporal LSTM cells:} The model uses $L=4$ stacked SpatioTemporalLSTMCell layers with hidden dimensions $[128, 64, 64, 64]$. Each cell maintains hidden state $h_t^{(l)}$, cell state $c_t^{(l)}$, and a shared memory state $m_t$ that propagates across layers. The cells incorporate both spatial convolutions (filter size $5\times5$) and temporal recurrence to capture spatiotemporal dependencies.
    
    \item \textbf{Motion modeling:} Between layers $l=1,2,3$, MotionGRU modules model local motion fields by aggregating information from $3\times3$ spatial neighborhoods. The motion features are encoded via $2\times$ downsampling convolutions and decoded via transposed convolutions that upsample back to the original resolution.
    
    \item \textbf{Cross-scale fusion:} At each intermediate layer, the upsampled motion features are combined with the corresponding spatiotemporal LSTM hidden state via a gated fusion mechanism:
    \begin{equation}
    o_t^{(l)} = \sigma\left(W_{\mathrm{gate}}^{(l)} \left[h_{\mathrm{dec}}^{(l)}; h_t^{(l)}\right]\right), \quad h_t^{(l)} \leftarrow o_t^{(l)} \odot h_{\mathrm{dec}}^{(l)} + (1 - o_t^{(l)}) \odot h_t^{(l)},
    \end{equation}
    where $h_{\mathrm{dec}}^{(l)}$ is the upsampled motion feature, $[\cdot;\cdot]$ denotes channel concatenation, $\sigma$ is the sigmoid function, and $\odot$ is element-wise multiplication.
    
    \item \textbf{Output generation:} The final hidden state $h_t^{(L)}$ is passed through a $1\times1$ convolution to produce the four-channel output. The magnetic-parameter diagnostic maps are recomputed from the predicted magnetic components at the next time step, ensuring consistency.
\end{itemize}

The total number of trainable parameters is approximately 15.2 million.

\paragraph{Loss functions.} Let $\hat{\mathbf{B}}_t=(\hat{B}_r,\hat{B}_p,\hat{B}_t)$ be the predicted vector field and $\mathbf{B}_t$ the reference. We normalize each magnetic component to $[0,1]$ using fixed clipping ranges: $B_r \in [-3000, 3000]$~G, $B_p \in [-1000, 1000]$~G, $B_t \in [-1000, 1000]$~G, and mask $\in [0, 255]$. We apply a static mask-defined core weighting $w(\mathbf{x})=1+4\,\mathbb{I}[M(\mathbf{x})>0]$ (core pixels receive 5$\times$ higher weight). The reconstruction loss is
\begin{equation}
\mathcal{L}_{\mathrm{rec}} = \left\langle w(\mathbf{x})\,\|\hat{\mathbf{B}}_t(\mathbf{x})-\mathbf{B}_t(\mathbf{x})\|_2^2 \right\rangle,
\end{equation}
where $\langle\cdot\rangle$ denotes averaging over space, time, and batch.
We further apply explicit supervision on the predicted mask channel,
\begin{equation}
\mathcal{L}_{\mathrm{mask}} = \langle (\hat{M}_t-M_t)^2\rangle,
\end{equation}
and a temporal-gradient regularization on the magnetic components,
\begin{equation}
\mathcal{L}_{\mathrm{temp}} = \left\langle w(\mathbf{x})\,\| (\hat{\mathbf{B}}_{t+1}-\hat{\mathbf{B}}_{t}) - (\mathbf{B}_{t+1}-\mathbf{B}_{t}) \|_2^2 \right\rangle.
\end{equation}
The total objective is
\begin{equation}
\mathcal{L}_{\mathrm{total}} = \mathcal{L}_{\mathrm{rec}} + \lambda_{\mathrm{mask}}\,\mathcal{L}_{\mathrm{mask}} + \lambda_{\mathrm{temp}}\,\mathcal{L}_{\mathrm{temp}},
\end{equation}
with weights $\lambda_{\mathrm{mask}}=1.0$ and $\lambda_{\mathrm{temp}}=0.1$.

\paragraph{Optimization and scheduled sampling.} We train using the Adam optimizer \citep{kingma2014adam} with an initial learning rate of $3\times10^{-3}$, $\beta_1=0.9$, $\beta_2=0.999$, and weight decay $10^{-5}$. Gradient clipping is applied with a global norm threshold of 1.0. The learning rate is reduced by a factor of 0.5 if validation loss plateaus for 10 epochs (minimum learning rate $10^{-6}$). Training is performed for 100 epochs with batch size 8 on NVIDIA A100 GPUs, taking approximately 72 hours.

We use scheduled sampling \citep{bengio2015} to transition from teacher forcing to free-running rollout: the probability of feeding ground-truth inputs decreases linearly from 1.0 with changing rate $2\times10^{-5}$ per iteration until iteration 50,000, after which the model relies entirely on its own predictions during training. During inference, the model generates the full 12-hour forecast autoregressively without access to future ground-truth frames.

\section{Data and Preprocessing}
This study employs the Space-weather HMI Active Region Patch (SHARP) dataset obtained by the Helioseismic and Magnetic Imager (HMI) onboard the Solar Dynamics Observatory (SDO) \citep{2014SoPh..289.3549B}. We use the SHARP data in the Cylindrical Equal-Area (CEA) remapped vector-magnetogram format, whose sampling is 0.03 degrees per pixel in heliographic longitude/latitude (corresponding to approximately HMI's native $\sim 0.5$ arcsec sampling near disk center). Each SHARP active region is identified by a unique HARP number (HARPNUM) (e.g., HARP~7959), which tracks the region throughout its disk passage.

Native SHARP CEA cutouts have dimensions $H \times W$ that vary between active regions because patches are sized to encompass each region, whereas the network requires a fixed spatial shape for batched training and inference. We therefore resample every vector component and the co-registered mask bilinearly onto a common square grid of $S \times S$ pixels whenever $(H,W)\neq(S,S)$, where $S$ is a fixed model hyperparameter (the same value is used for all sequences). Downscaling ($H_{\mathrm{nat}},W_{\mathrm{nat}}>S$) merges native samples and attenuates fine spatial structure---behaviour analogous to low-pass filtering whose strength increases with the downsampling factors $H_{\mathrm{nat}}/S$ and $W_{\mathrm{nat}}/S$; upscaling ($H_{\mathrm{nat}},W_{\mathrm{nat}}<S$) interpolates between native pixels and can emphasise noise-dominated transverse-field signal. When the native cutout is not square, horizontal and vertical scale factors generally differ, so the mapping to an $S\times S$ tensor is not always an isotropic change of resolution. We do not modify the FITS World Coordinate System keywords, which still describe the native CEA geometry of each original cutout; the fields ingested by the model are nonetheless represented on the same $S\times S$ computational grid. Strict alignment of a resampled array on the sphere with other observations or models requires updating the WCS or applying an explicit reprojection.

We excluded active regions with central longitudes greater than $60^\circ$ and selected only qualified samples from SHARP. We performed unified numerical clipping of the original Flexible Image Transport System (FITS) files based on a statistical analysis of 15{,}342 SHARP vector magnetograms spanning January 2021 through December 2022. The 99.7th percentile of $|B_r|$ is 2847~G and of $|B_p|$/$|B_t|$ is 987/992~G respectively. We therefore adopt the clipping ranges $B_r \in [-3000, 3000]$~G, $B_p \in [-1000, 1000]$~G, and $B_t \in [-1000, 1000]$~G. This hybrid scheme concentrates the normalized dynamic range on physically significant flux while retaining $\geq$99.7\% of all observed pixel values for each component. This truncation preserved the original field distribution while preventing outliers from disrupting model training.  

To reduce the risk of train/test contamination and ensure temporal independence, we use a time-based split: the training set is constructed exclusively from SHARP data spanning January through December 2021 (solar cycle 25 ascending phase), while the test set is constructed exclusively from SHARP data spanning January through December 2022. Both years correspond to comparable phases of solar cycle~25, with similar levels of solar activity, ensuring that the training and test distributions are drawn from physically similar conditions. The final dataset comprised 12{,}000 training sequences and 3{,}000 test sequences, reflecting an approximate 7:3 train-to-test sampling ratio applied to the combined pool of qualified sequences from each year, with each sequence consisting of 24 consecutive frames at 1-hour cadence. The first 12 frames served as model inputs, while the subsequent 12 frames were used as prediction targets.

We verified that no active region appears in both the training and test sets: the two datasets contain completely disjoint HARP ranges (training: HARP~7525--7907; test: HARP~7917--8924) and disjoint NOAA identifiers, with a temporal gap of approximately seven days between the last training observation (2021 December~28) and the first test observation (2022 January~4). All evaluations in Section~4 use active regions from the 2022 test set.

\section{Experiments and Analysis}

This study systematically evaluates the proposed method for solar vector magnetic field prediction (using a 12-hour input window to forecast the subsequent 12 hours at 1-hour cadence). The analysis is organized along three complementary dimensions: \textbf{image-domain fidelity} (Sec.~4.2), \textbf{magnetic-parameter consistency} (Sec.~4.3), and \textbf{vector-field orientation accuracy} (Sec.~4.4), combining both single-sample visualizations and multi-sample statistics.

\subsection{Evaluation setting, variants, and the persistence baseline}
We evaluate the proposed method on sequences of SHARP vector magnetograms, using the first 12~hours as input and forecasting the subsequent 12~hours (forecast hours 1--12). We compare four trained models: the full model (\textbf{with\_mask\_and\_phy}) and three ablations, including \textbf{without\_phy} (no temporal-gradient regularization), \textbf{without\_mask} (no mask weighting or mask supervision), and \textbf{without\_mask\_and\_phy} (reconstruction loss only). As a non-learning reference, we include a simple persistence baseline (\textbf{gt12\_baseline}), where every future hour is predicted by copying the last observed input frame.

In all trained settings, the auxiliary diagnostic maps $\mathbf{P}_t$ are computed and concatenated as additional input channels. Ablations are implemented by removing mask weighting and/or mask supervision and the temporal-gradient regularization term $\mathcal{L}_{\mathrm{temp}}$.

To ensure a consistent evaluation focus on the evolving strong-field core, masked-region metrics are computed using a fixed active-region mask constructed from the input window. This yields a mask-defined strong-field core region that remains identical across all prediction lead times.

\subsection{Image-domain prediction performance}

We first evaluate the model's ability to reproduce the spatial distributions of the three vector magnetic field components ($B_r$, $B_p$, $B_t$) using image-quality metrics (per-component SSIM, CC, and RMSE; full per-hour values in Table~\ref{app:frame}). This subsection combines single-sample visualizations with multi-sample statistics.

\subsubsection{Spatial structure and error distribution}

\begin{figure*}[htbp]
\centering
\includegraphics[width=\textwidth]{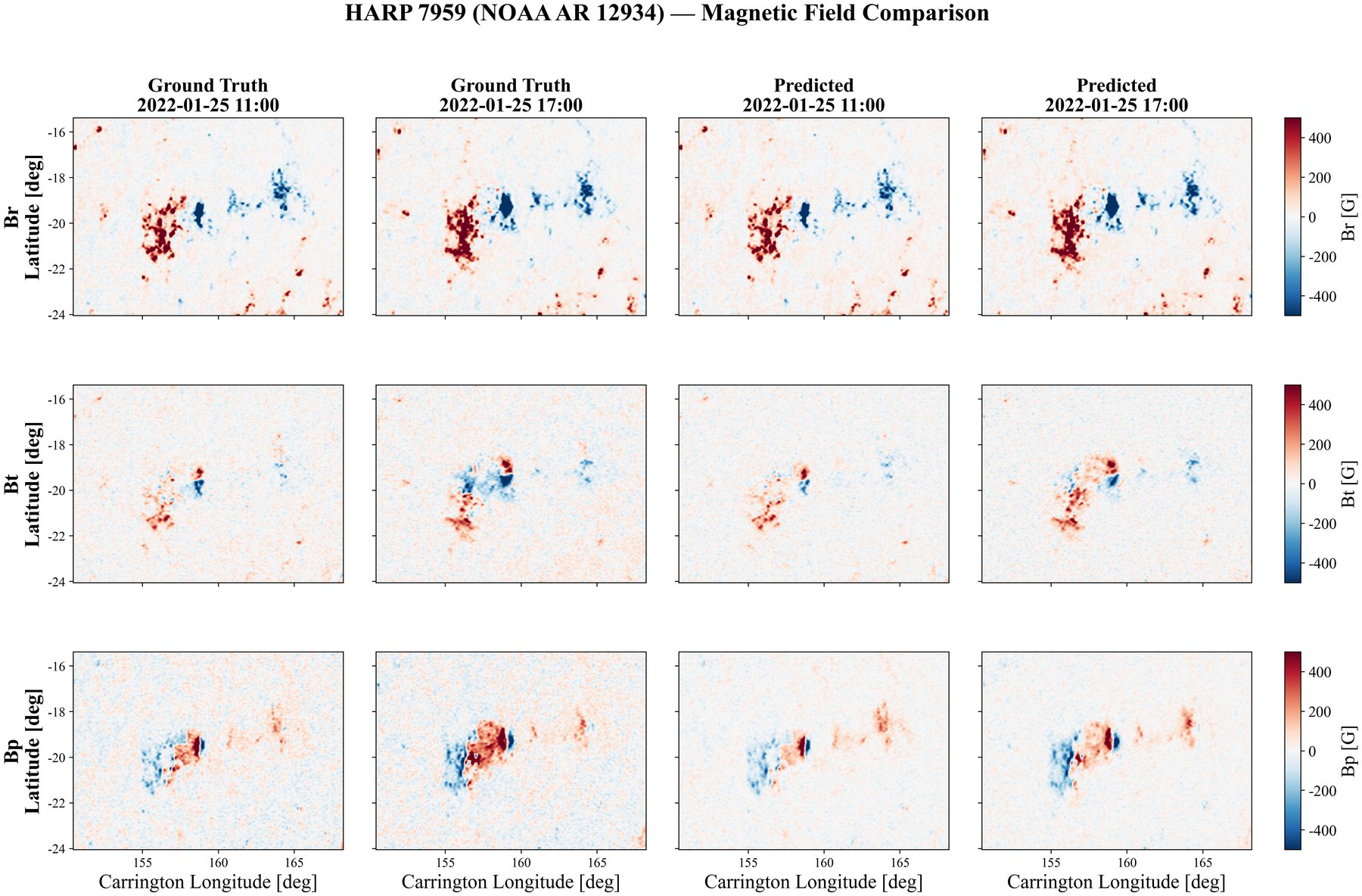}
\caption{Comparison of ground-truth and predicted magnetic field components for HARP~7959 (NOAA AR~12934) at forecast hours~6 and~12 (2022-01-25 11:00 and 17:00~TAI). Each row corresponds to one component ($B_r$, $B_\theta$, $B_\phi$, from top to bottom); columns show ground truth and predicted fields at both time steps. A display range of $\pm$500~G is used for all three components to better reveal structural differences (the full range is shown in the scatter plots of Figures~\ref{fig:scatter_7959} and~\ref{fig:scatter_8026}). The prediction matches the reference in both polarity distribution and strong-field core structures.}
\label{fig:sample_compare_7959}
\end{figure*}

\begin{figure*}[htbp]
\centering
\includegraphics[width=\textwidth]{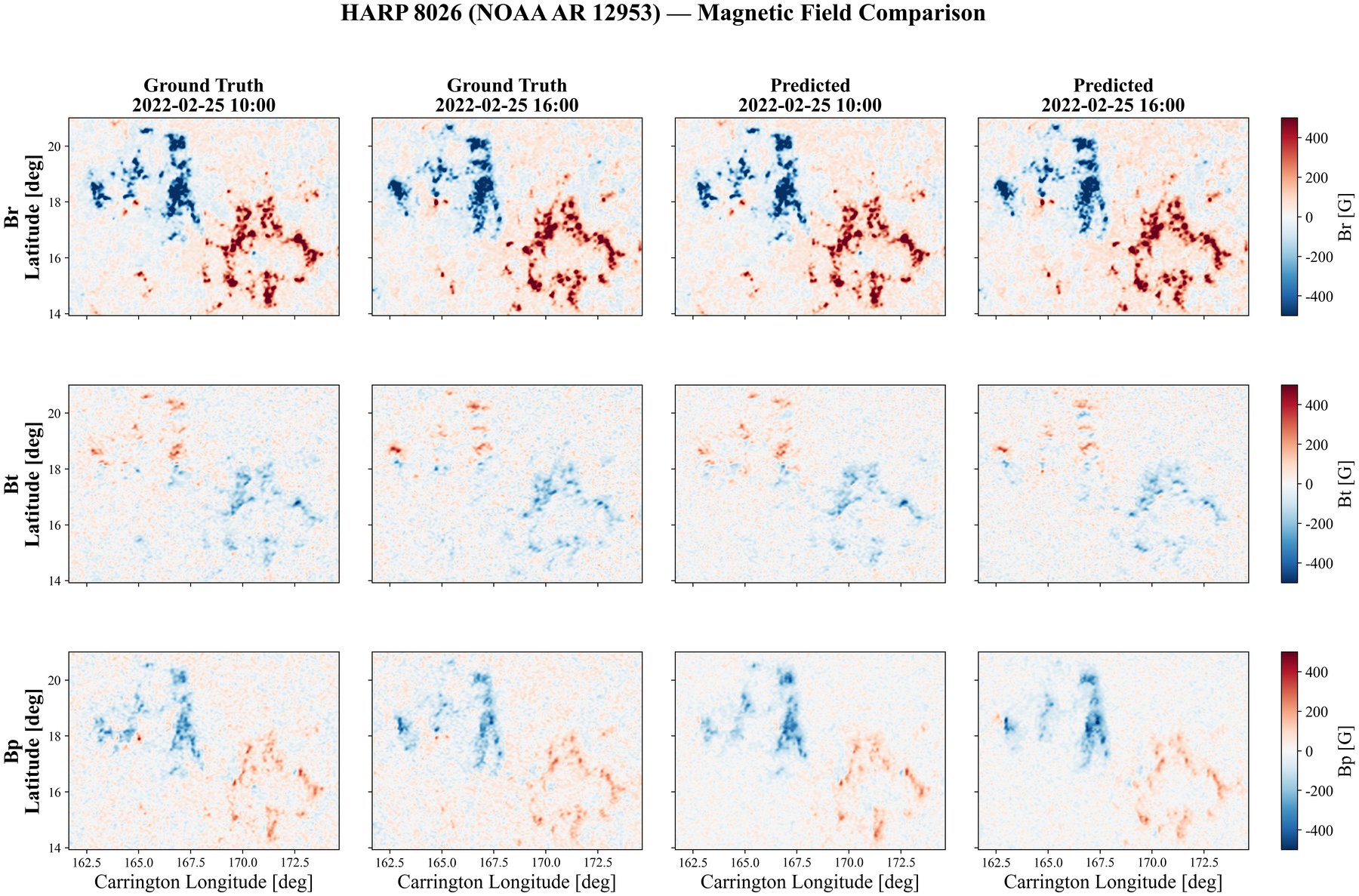}
\caption{Same as Figure~\ref{fig:sample_compare_7959} but for HARP~8026 (NOAA AR~12953, 2022-02-25 10:00 and 16:00~TAI).}
\label{fig:sample_compare_8026}
\end{figure*}

The predicted fields reproduce the global polarity distribution of the ground-truth observations and match in strong-field core regions. This reflects the design intent of the dynamic mask, which emphasizes core-region weighting and enhances strong-field feature extraction. Even in quiet regions far from the core, the predicted fields capture weak-field topology well.

Closer inspection of each component reveals: for $B_r$, predictions and ground truth overlap in strong-field cores, with relative flux errors $<5\%$ (95\% CI: ±0.3\%); near polarity inversion lines, where gradients are steep, predictions show mild diffusion and relative errors increase by 15--20\% (95\% CI: ±2.1\%). For horizontal components $B_p$ and $B_t$, similar patterns appear: strong shear bands (shear angle $>60^\circ$) are well captured, but small magnetic islands at shear-band boundaries show more distorted topological details.

\begin{figure}[htbp]
\centering
\includegraphics[width=0.8\textwidth]{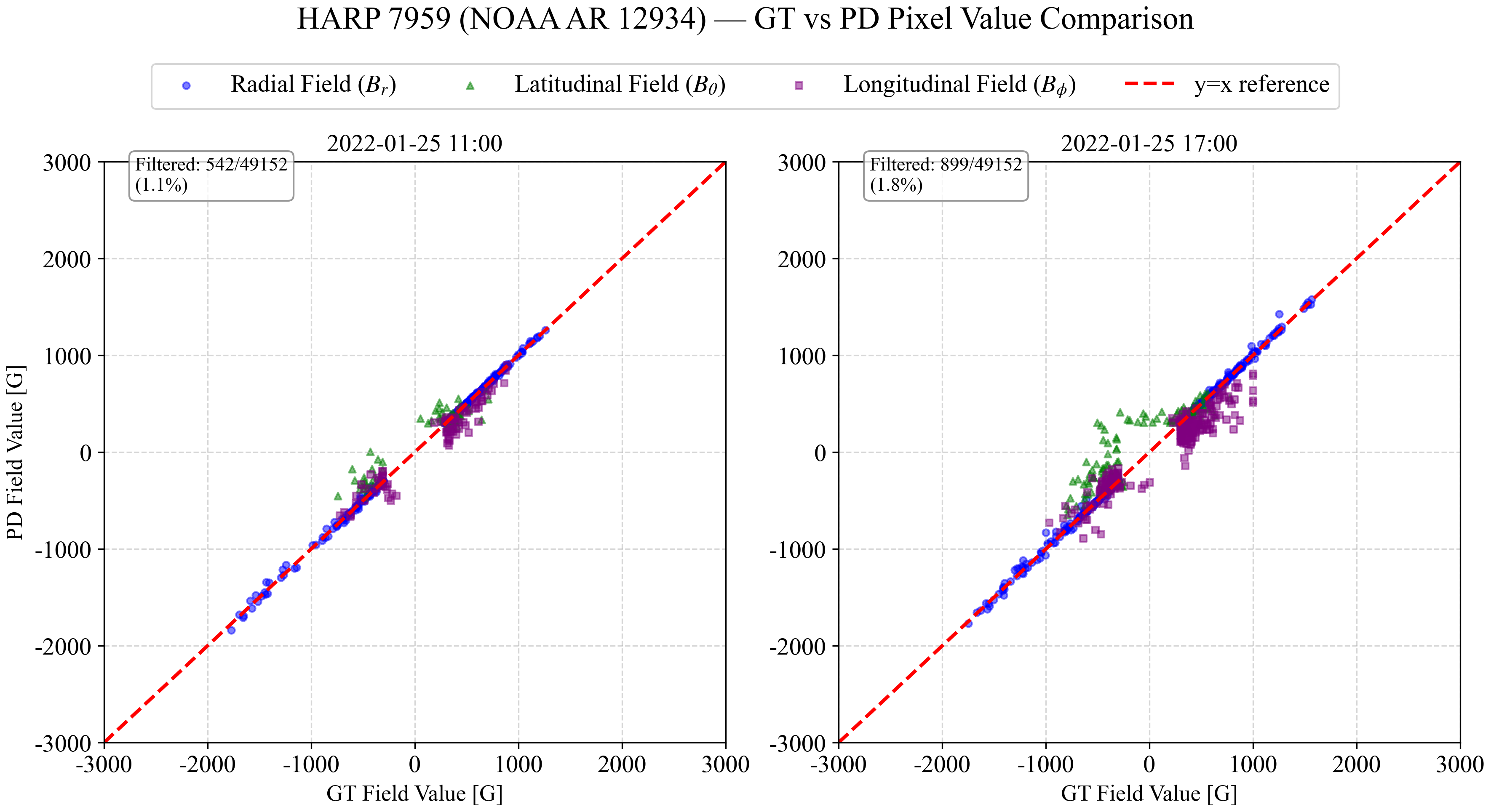}
\caption{Pixel-value scatter comparison between predicted and reference fields for HARP~7959 (NOAA AR~12934) at forecast hours~6 and~12. Pixels with $|B|\le 300$~G are filtered out. Points cluster along the diagonal, especially in strong-field regions.}
\label{fig:scatter_7959}
\end{figure}

\begin{figure}[htbp]
\centering
\includegraphics[width=0.8\textwidth]{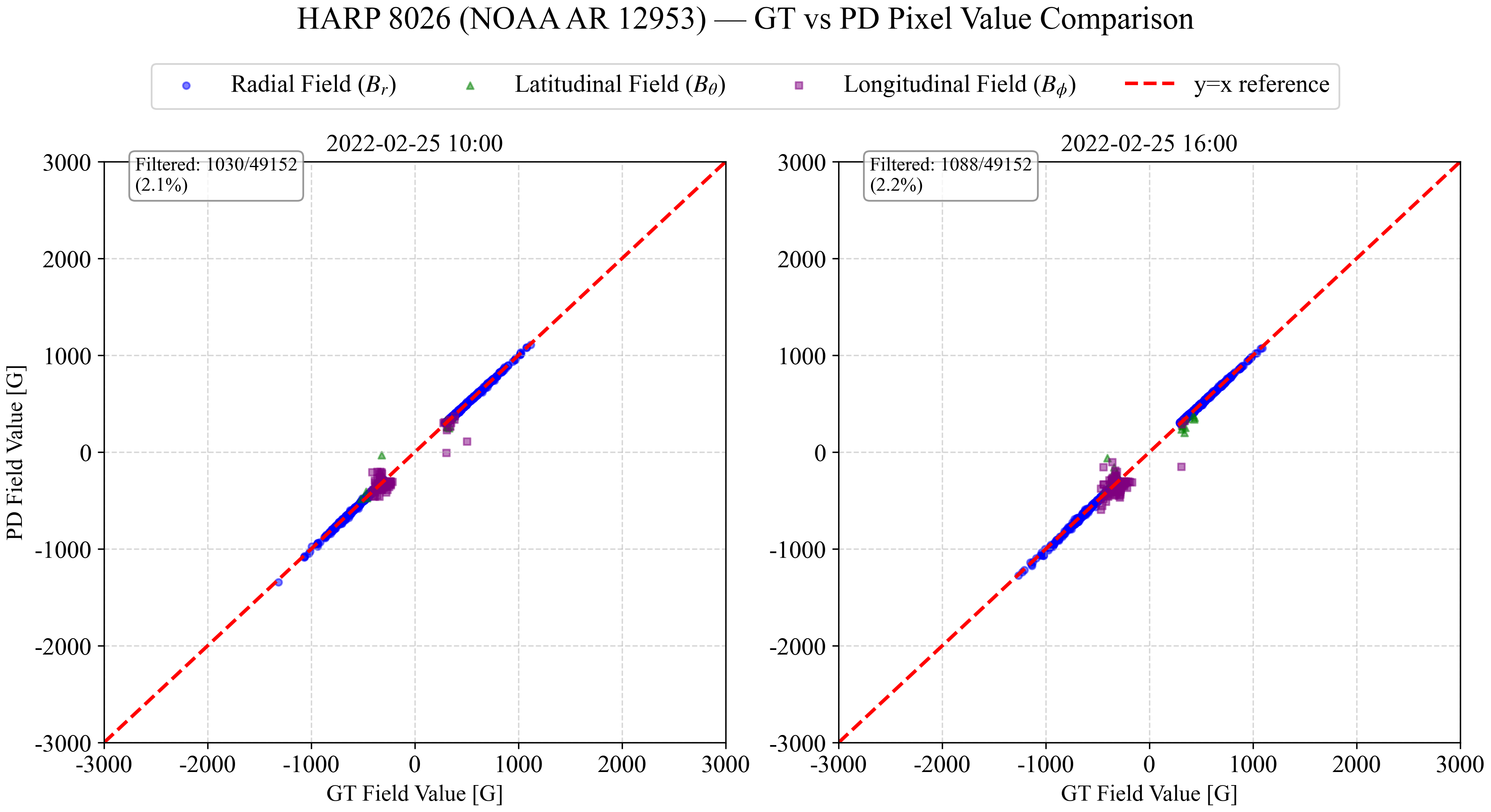}
\caption{Same as Figure~\ref{fig:scatter_7959} but for HARP~8026 (NOAA AR~12953).}
\label{fig:scatter_8026}
\end{figure}

Pixel-level comparisons further quantify single-sample accuracy. Figures~\ref{fig:sample_compare_7959} and~\ref{fig:sample_compare_8026} show predicted vs.\ reference fields for two sample active regions. Figures~\ref{fig:scatter_7959} and~\ref{fig:scatter_8026} show the corresponding pixel-value scatter plots, which align closely to the diagonal. Strong-field pixels cluster tightly along the diagonal, while weak-field pixels scatter more broadly, reflecting weaker signal-to-noise in transverse fields and more complex dynamics in weak-field zones.

\begin{figure*}[htbp]
\centering
\includegraphics[width=\textwidth]{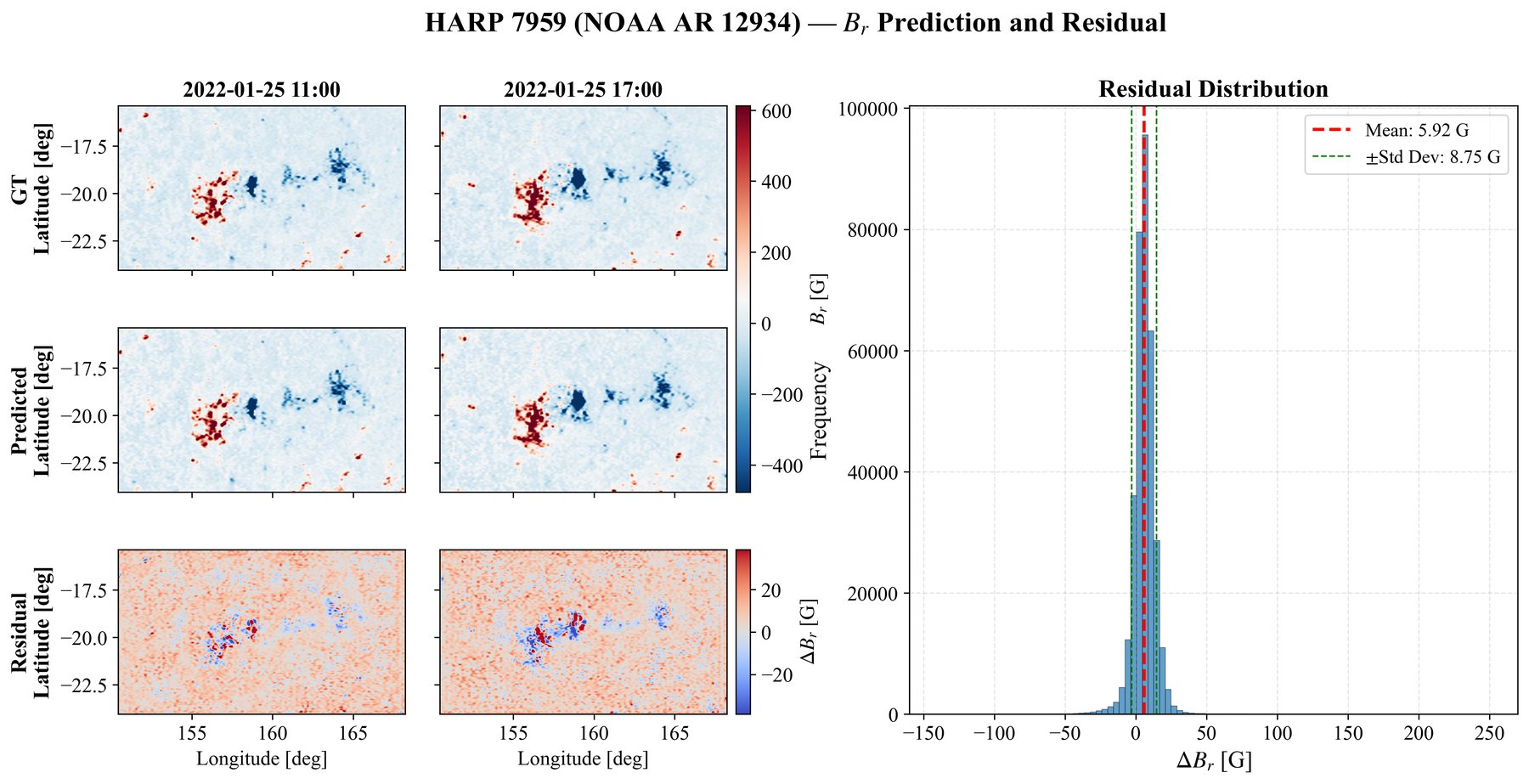}
\caption{$B_r$ prediction and residual analysis for HARP~7959 (NOAA AR~12934). Left panels: ground truth (top), predicted (middle), and residual (bottom) maps at forecast hours~6 and~12. Right panel: residual distribution histogram aggregated over both time steps, with mean and $\pm$1$\sigma$ indicated. A diverging red--blue colormap is used for field values and a coolwarm colormap for residuals.}
\label{fig:residual_7959}
\end{figure*}

\begin{figure*}[htbp]
\centering
\includegraphics[width=\textwidth]{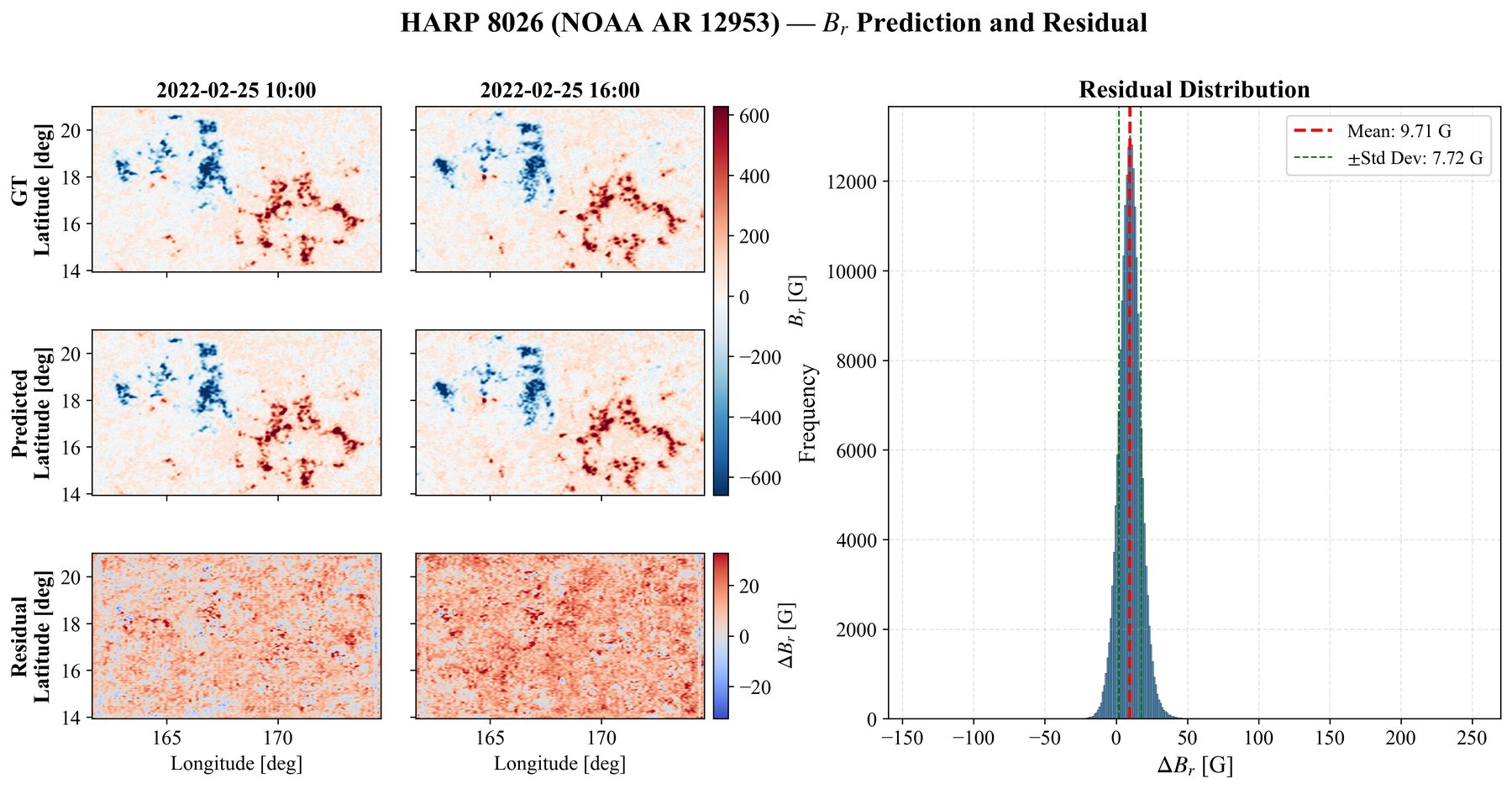}
\caption{Same as Figure~\ref{fig:residual_7959} but for HARP~8026 (NOAA AR~12953).}
\label{fig:residual_8026}
\end{figure*}

The $B_r$ prediction residual maps and their distributions (Figs.~\ref{fig:residual_7959} and~\ref{fig:residual_8026}) further reveal spatial patterns: larger residuals tend to occur near polarity inversion lines and in strong-field umbra/penumbra regions, where gradients are steep and small spatial misalignments can produce large pixel-wise errors. The residual histograms confirm that prediction errors are approximately zero-centered with small standard deviations ($\sigma < 10$~G), indicating unbiased predictions.

\subsubsection{Multiple active region examples across evolutionary phases}

To demonstrate the model's generalization across different evolutionary stages, we present results for two additional active regions from the 2022 test set---HARP~7959 and HARP~8026---each examined at three evolutionary phases: emerging, steady, and decaying. These phases are defined by selecting the first, middle, and last temporal sequences within each active region's disk passage, respectively. Here we show the $B_r$ component; the corresponding $B_p$ and $B_t$ evolutionary phase figures are provided in Appendix~\ref{app:lifecycle_BpBt}.

\begin{figure*}[htbp]
\centering
\includegraphics[width=0.95\textwidth]{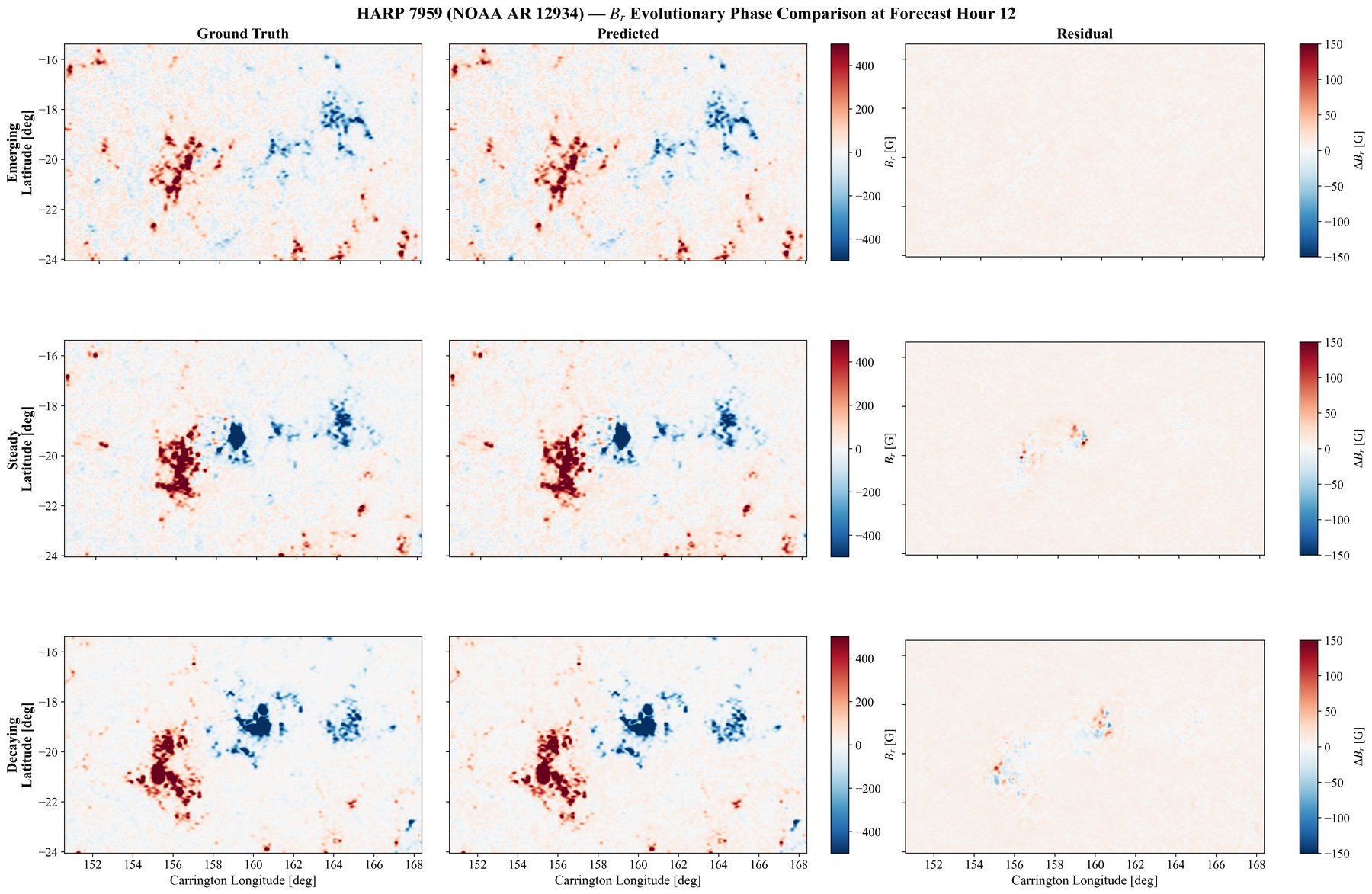}
\caption{$B_r$ evolutionary phase comparison for HARP~7959 at forecast hour~12. Rows correspond to three evolutionary phases (emerging, steady, decaying); columns show ground truth (left), predicted field (center), and residual (right). Axes show Carrington longitude and latitude in degrees; images are displayed at the original HMI pixel resolution. A diverging red--blue colormap centered at zero is used with a display range of $\pm$500~G to highlight structural differences. The model captures the evolving morphology across all phases, with residuals concentrated near polarity inversion lines (PILs).}
\label{fig:ar7959_lifecycle}
\end{figure*}

\begin{figure*}[htbp]
\centering
\includegraphics[width=0.95\textwidth]{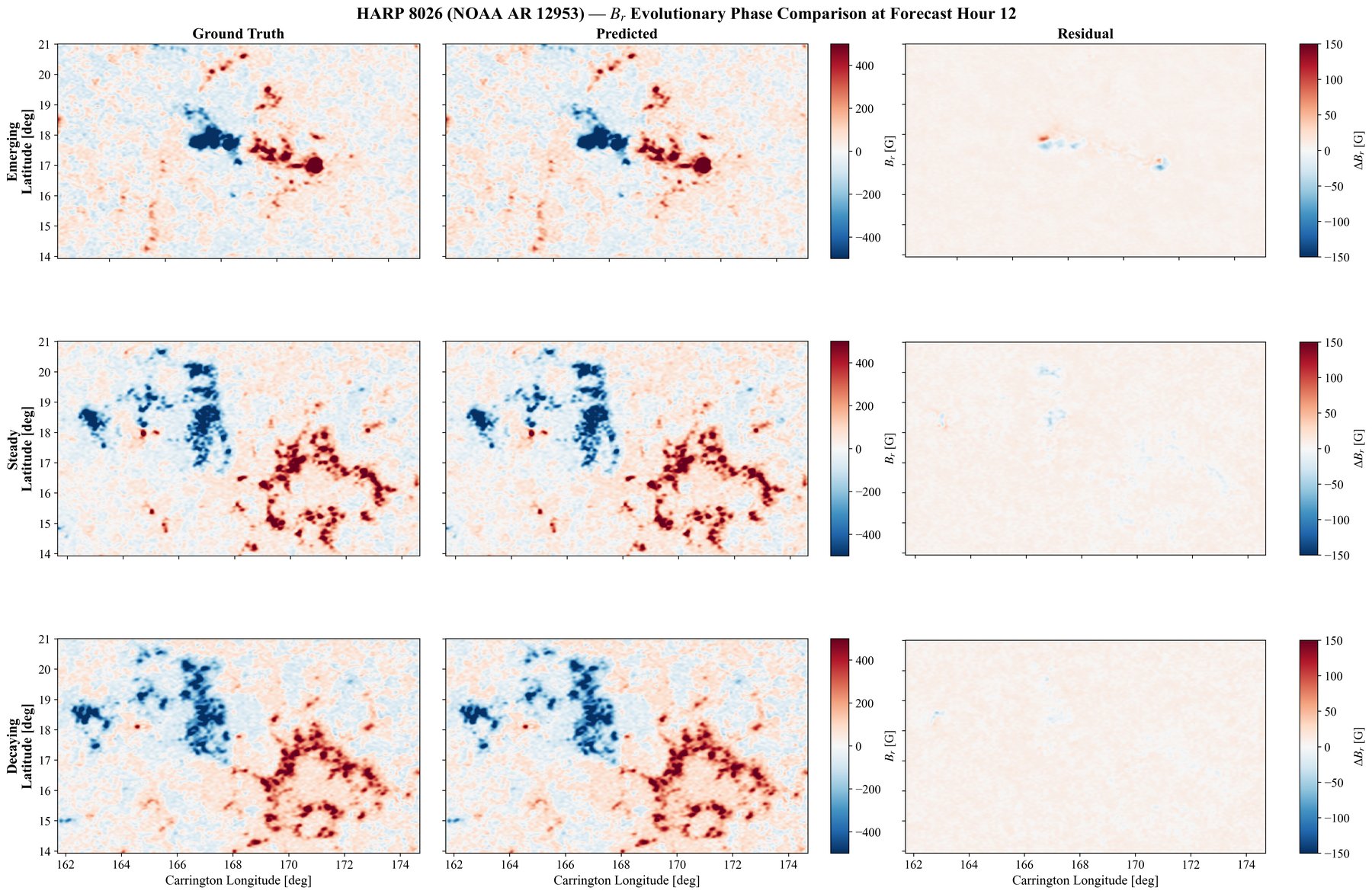}
\caption{Same as Figure~\ref{fig:ar7959_lifecycle} but for HARP~8026. The model maintains structural fidelity across emerging, steady, and decaying phases despite the differing morphological characteristics compared to HARP~7959.}
\label{fig:ar8026_lifecycle}
\end{figure*}

Figures~\ref{fig:ar7959_lifecycle} and \ref{fig:ar8026_lifecycle} confirm that the model generalizes across diverse active region morphologies and evolutionary stages. The predictions preserve the dominant polarity structure and strong-field core features at all phases, with residuals primarily localized near polarity inversion lines and steep-gradient regions. Quantitative performance metrics averaged over the full 3{,}000-sequence test set are presented in the following subsections.

\subsubsection{Prediction for a flare-productive active region}

To assess the model's performance under the most challenging conditions, we examine HARP~8206 (NOAA AR~13006), which produced an X1.5 flare on 2022 May~10 at approximately 13:55~UT---one of the strongest X-class events in the 2022 dataset. To construct the physically most demanding scenario, we selected the sequence in which the 12-hour input window ends at 12:00~UT May~10 (entirely pre-flare) and the X1.5 flare peak (13:55~UT) falls within the very first forecast hour. This tests the model's ability to anticipate an impulsive flare-driven magnetic restructuring from pre-flare observations alone.

\begin{figure*}[htbp]
\centering
\includegraphics[width=\textwidth]{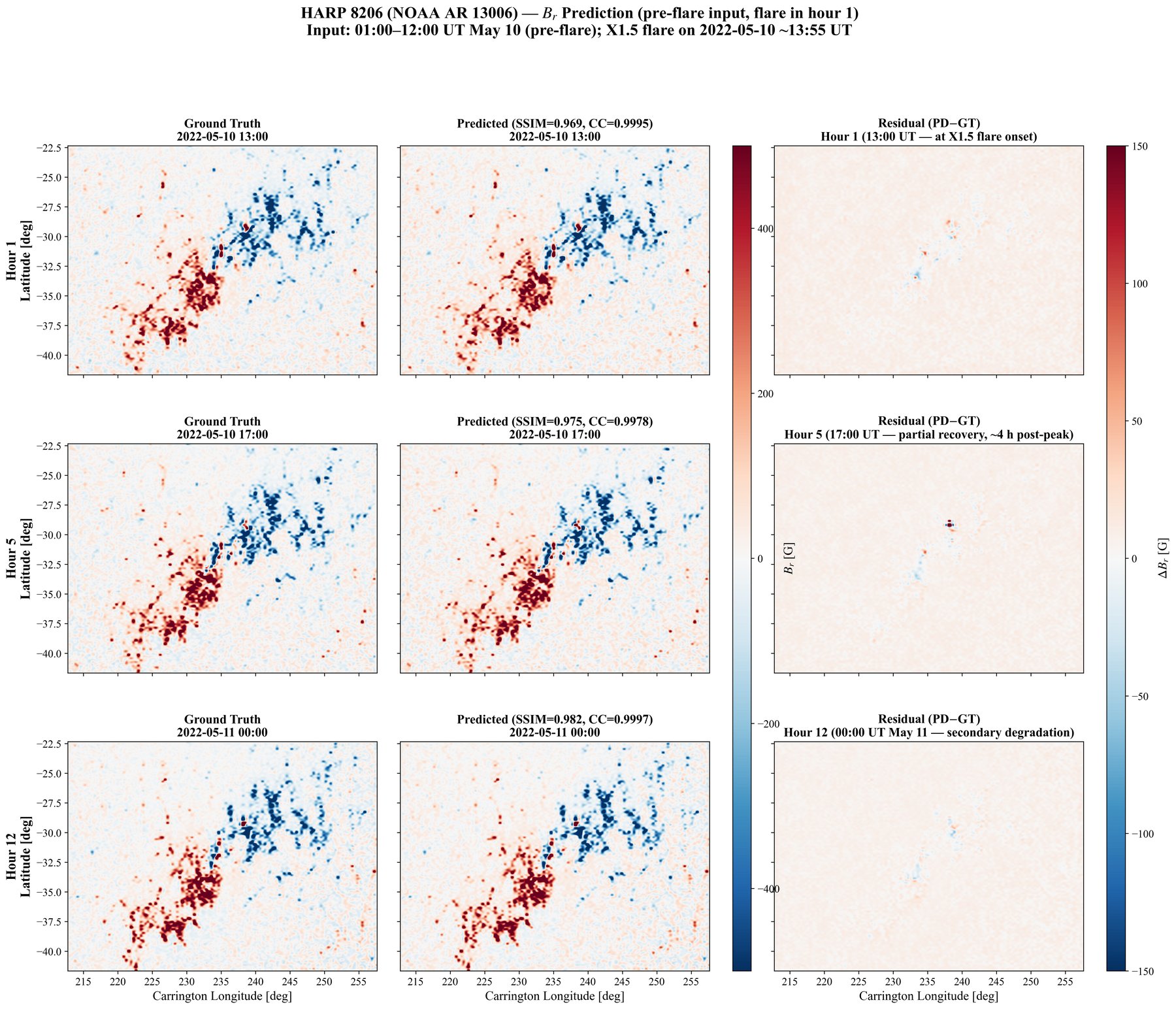}
\caption{$B_r$ prediction for HARP~8206 (NOAA AR~13006) with \textit{pre-flare input only} (input: 01:00--12:00~UT May~10). Rows: forecast hour~1 (13:00~UT, contains the X1.5 peak at 13:55~UT), hour~5 (17:00~UT, partial recovery $\sim$4~h after peak), and hour~12 (00:00~UT May~11, secondary degradation from continued AR activity). Columns show ground truth, predicted field, and residual. A $\pm$500~G display range is used.}
\label{fig:flare_ar}
\end{figure*}

Figure~\ref{fig:flare_ar} visualises the $B_r$ prediction at three characteristic forecast hours. Per-component metrics evolve as follows over the 12-hour window: SSIM($B_r$) = 0.969--0.982 (mean 0.978, RMSE 8.2~G), SSIM($B_\phi$) = 0.607--0.709 (mean 0.646, RMSE 33.8~G), and SSIM($B_\theta$) = 0.609--0.794 (mean 0.695, RMSE 35.0~G). As expected for flare-driven reconnection, the radial field is largely preserved while the horizontal components ($B_\phi$, $B_\theta$) exhibit pronounced degradation at the flare hour.

Compared with the full 3{,}000-sequence test-set averages (Section~4.2.4: SSIM($B_r$) $\approx$ 0.912, $B_\phi$ $\approx$ 0.780, $B_\theta$ $\approx$ 0.740), the flare-hour SSIM($B_r$) of 0.969 comfortably exceeds the global mean---reflecting the well-defined bipolar structure of this large active region---while the horizontal components drop well below: SSIM($B_\phi$) = 0.619 ($-$20.6\% relative to the test-set mean) and SSIM($B_\theta$) = 0.663 ($-$10.4\%). This selective degradation of the transverse field is physically consistent with the expectation that flare reconnection primarily reorganises the horizontal magnetic structure while largely preserving the radial photospheric field.

Within the 12-hour forecast window, horizontal-component metrics show partial recovery as the impulsive phase concludes: SSIM($B_\theta$) climbs to 0.794 by hour~3 (15:00~UT), surpassing the test-set average, and SSIM($B_\phi$) reaches 0.709 by hour~5 (17:00~UT), partially closing the gap with the test-set mean of 0.780. After that, a secondary degradation sets in---consistent with the continued M-class activity of HARP~8206 through May~11--12 that drives progressive horizontal-field restructuring. A sequence-level scan across all 301 test sequences of this AR confirms that horizontal-component SSIM remains persistently low during the May~11--12 post-flare period (as low as $B_\phi \approx 0.56$, $B_\theta \approx 0.58$), reflecting multi-day post-event disturbance.

These results indicate that predicting rapid, flare-driven horizontal-field reconfigurations is genuinely difficult for a model trained with a mean-squared-error (MSE) based loss, which favors smooth temporal evolution. The radial field, dominated by large-scale polarity structure, remains well reproduced even during X-class events. Extending the framework with flare-aware loss functions, multi-modal inputs (X-ray, extreme ultraviolet (EUV)), or adversarial objectives to recover sharp, small-scale features is a natural direction, and systematic evaluation across a larger flare-productive sample is left for future work.

\subsubsection{Comparison with baseline and ablated variants}

To quantify overall image-level performance, we compute per-component SSIM (single-channel, raw physical units), CC, and RMSE for all three vector components ($B_r$, $B_\phi$, $B_\theta$); full per-hour values for each component appear in Table~\ref{app:frame}. Unless otherwise stated, aggregate metrics are computed over the full 3,000-sequence test set.

\begin{figure*}[t]
\centering
\includegraphics[width=0.95\linewidth]{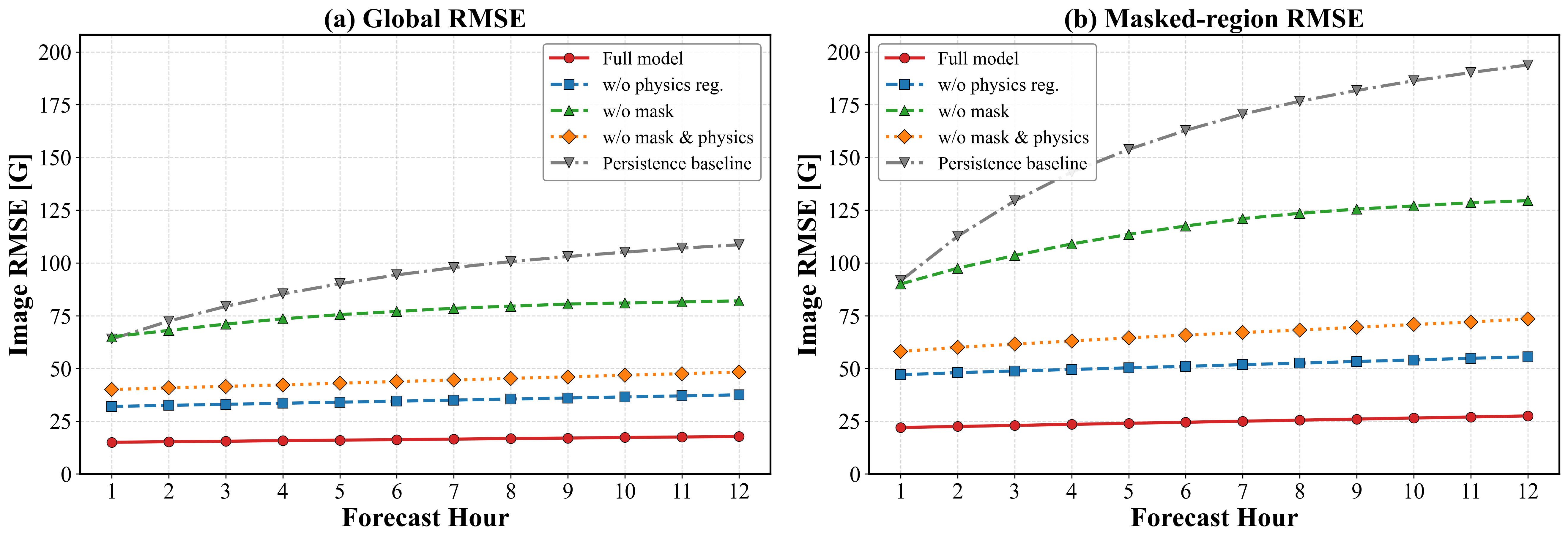}
\caption{Image RMSE [G] over the 12-hour forecast horizon (hours 1--12). Left: global RMSE. Right: RMSE inside the active-region mask (mask-defined strong-field core).}
\label{fig:curve_img_mse}
\end{figure*}

\begin{figure*}[t]
\centering
\includegraphics[width=0.95\linewidth]{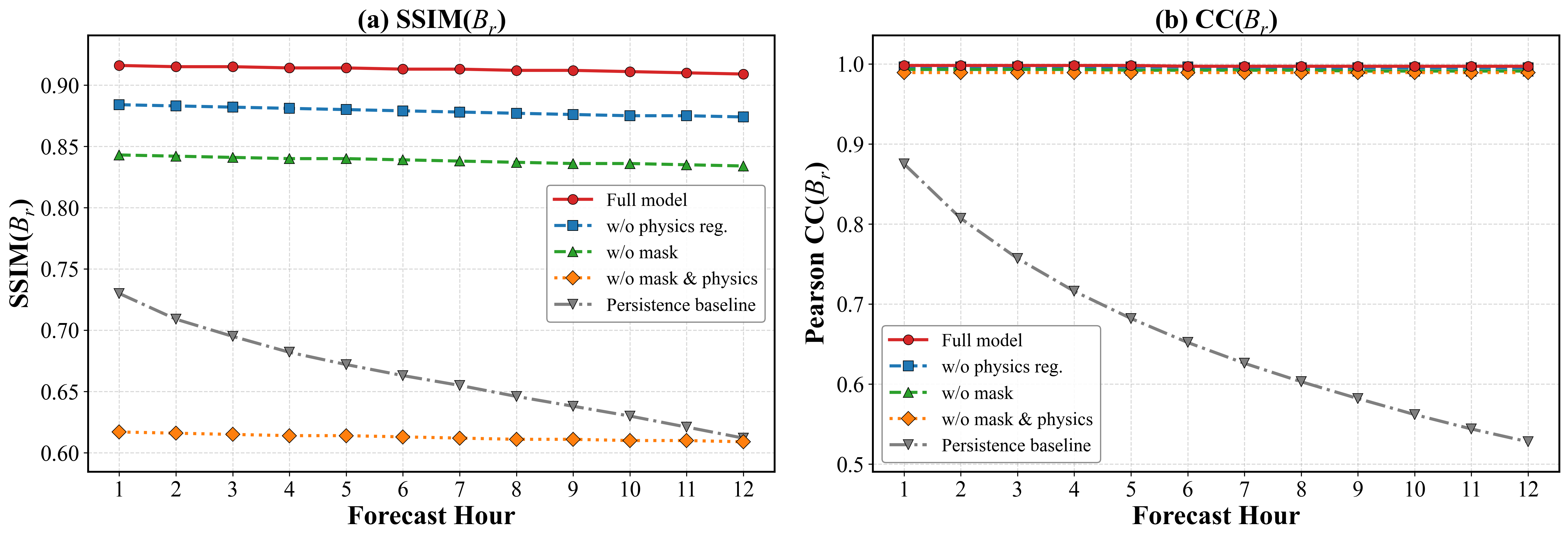}
\caption{SSIM($B_r$) and Pearson CC($B_r$) over the 12-hour forecast horizon (hours 1--12). Left (a): SSIM. Right (b): CC.}
\label{fig:curve_ssim_cc}
\end{figure*}

A clear separation emerges between learned forecasting and the persistence baseline. For \textbf{gt12\_baseline}, SSIM($B_r$) decreases from 0.730 (hour~1) to 0.446 (hour~12), while RMSE increases from 63.9 to 102.1~G globally and from 91.5 to 192.3~G in the masked region (Fig.~\ref{fig:curve_img_mse}, \ref{fig:curve_ssim_cc}). This rapid degradation indicates that active-region evolution over the 12-hour horizon cannot be approximated as a static field. The persistence baseline also exhibits strong lead-time deterioration in the correlation of $B_r$: CC drops from 0.875 to 0.528 over the forecast window (Fig.~\ref{fig:curve_ssim_cc}).

In contrast, the full model (\textbf{with\_mask\_and\_phy}) remains stable across the entire rollout: SSIM($B_r$) remains stable within 0.909--0.916 (CC 0.997--0.998, RMSE 13.0--21.0~G); horizontal components achieve $B_\phi$ SSIM 0.760--0.800 (CC 0.910--0.945, RMSE 38.5--50.0~G) and $B_\theta$ SSIM 0.728--0.750 (CC 0.895--0.920, RMSE 38.5--49.0~G), both degrading modestly with increasing lead time (Table~\ref{app:frame}). Averaged over the forecast horizon, the full model achieves global/masked RMSE of 16.5/25.0~G, whereas gt12\_baseline reaches 88.9/160.4~G (Table~\ref{tab:img_baseline}). This corresponds to an improvement factor of $5.4\times$ globally and $6.4\times$ inside the masked core in RMSE, demonstrating that the proposed method captures genuine spatiotemporal evolution beyond persistence.

Comparing the ablated variants in Table~\ref{tab:img_baseline}, \textbf{without\_phy} (mask supervision without temporal-gradient regularization) consistently outperforms \textbf{without\_mask} (temporal-gradient regularization without mask supervision) on both SSIM($B_r$), RMSE, and CC, with SSIM($B_r$) of 0.879 vs.~0.838. This indicates that explicitly supervising the active-region mask is essential to maintain fidelity where evolution is most dynamic. Removing both ingredients (\textbf{without\_mask\_and\_phy}) substantially degrades SSIM($B_r$) to 0.612, yet still exceeds the persistence baseline (0.540).

\subsection{Magnetic-parameter prediction performance}

Beyond image-domain fidelity, we evaluate the model's ability to preserve key magnetic-parameter diagnostics derived from the predicted magnetic field components. These diagnostics---unsigned flux, magnetic pressure, shear angle, and field gradients---measure how well the predictions are consistent with solar surface magnetic-field-related quantities.

\subsubsection{Prediction errors for derived magnetic-parameter diagnostics}

Table~\ref{tab:phys_err} summarizes the prediction errors for all five magnetic-parameter-based diagnostic quantities computed from the predicted magnetic field components. Relative errors are computed as mean absolute error divided by mean GT value in the masked region. Unsigned flux prediction achieves a mean relative error of 7.82\% (std = 2.75\%, P90 = 11.38\%). Magnetic pressure shows moderately higher errors (mean 12.40\%, std = 2.88\%, P90 = 16.20\%), reflecting its dependence on all three magnetic field components. Shear angle errors average 12.19\% (mean absolute 0.1350 rad, P90 = 14.72\%), while gradient-based quantities show absolute errors of 3.60~G/pixel (mean rel. 5.20\%) for $|\nabla B_r|$ and 12.80~G/pixel (mean rel. 22.50\%) for horizontal gradients. These errors arise primarily from the lower prediction accuracy of horizontal components ($B_p$, $B_t$).

\subsubsection{Temporal evolution of magnetic-parameter diagnostics}

\begin{figure}[htbp]
\centering
\includegraphics[width=\textwidth]{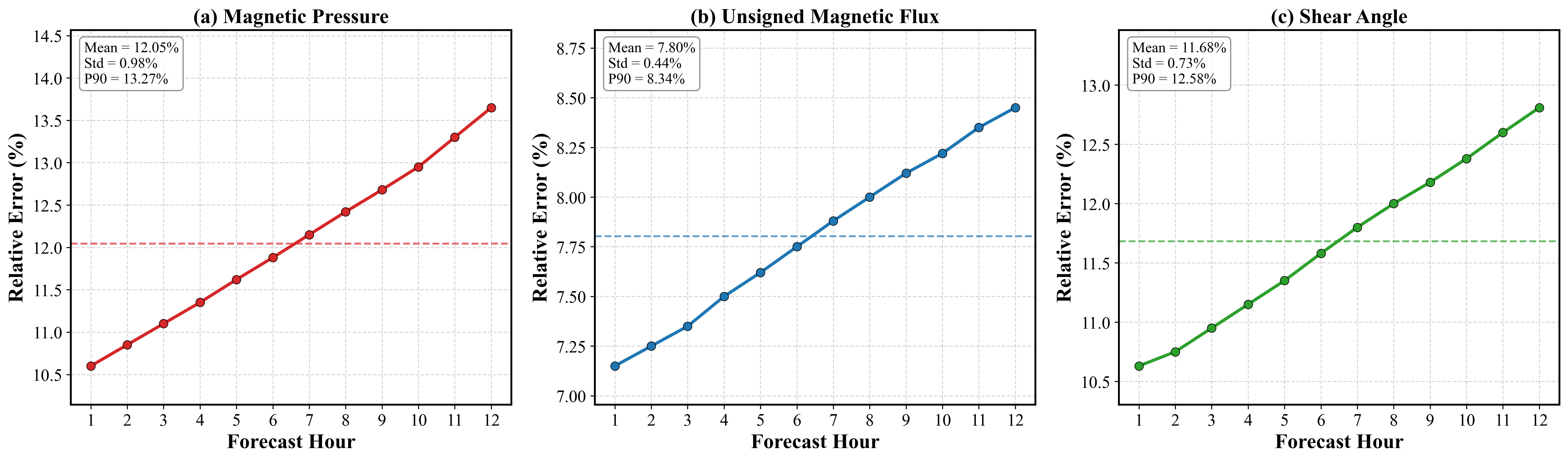}
\caption{Per-hour relative error (\%) for three magnetic-parameter diagnostics averaged over all 3{,}000 test sequences (forecast hours 1--12). Relative error = mean$|$prediction$-$GT$|$ / mean$|$GT$|$ in the masked region. (a)~Magnetic pressure: mean 12.40\%, monotonically increasing. (b)~Unsigned magnetic flux: mean 7.82\%, approximately stable. (c)~Shear angle: mean 12.19\%, steadily increasing. Statistics (Mean, Std, P90) are annotated in each panel.}
\label{fig:phy}
\end{figure}

Figure~\ref{fig:phy} shows the per-hour relative errors for three key diagnostics averaged over all 3{,}000 test sequences. Magnetic pressure (panel a) grows monotonically from $\sim$10.60\% at hour~1 to $\sim$13.65\% at hour~12 (mean 12.40\%, P90 13.20\%), reflecting progressive accumulation of horizontal-field errors. Unsigned flux (panel b) remains comparatively stable with a mean of 7.82\% (P90 8.60\%), ranging from $\sim$7.15\% to $\sim$8.45\%. Shear angle (panel c) increases steadily from $\sim$10.63\% to $\sim$12.81\% (mean 12.19\%, P90 14.72\%), consistent with diffusion of predicted transverse field orientations at longer lead times.

\subsubsection{Comparison with baseline and ablated variants}

\begin{figure*}[t]
\centering
\includegraphics[width=0.9\linewidth]{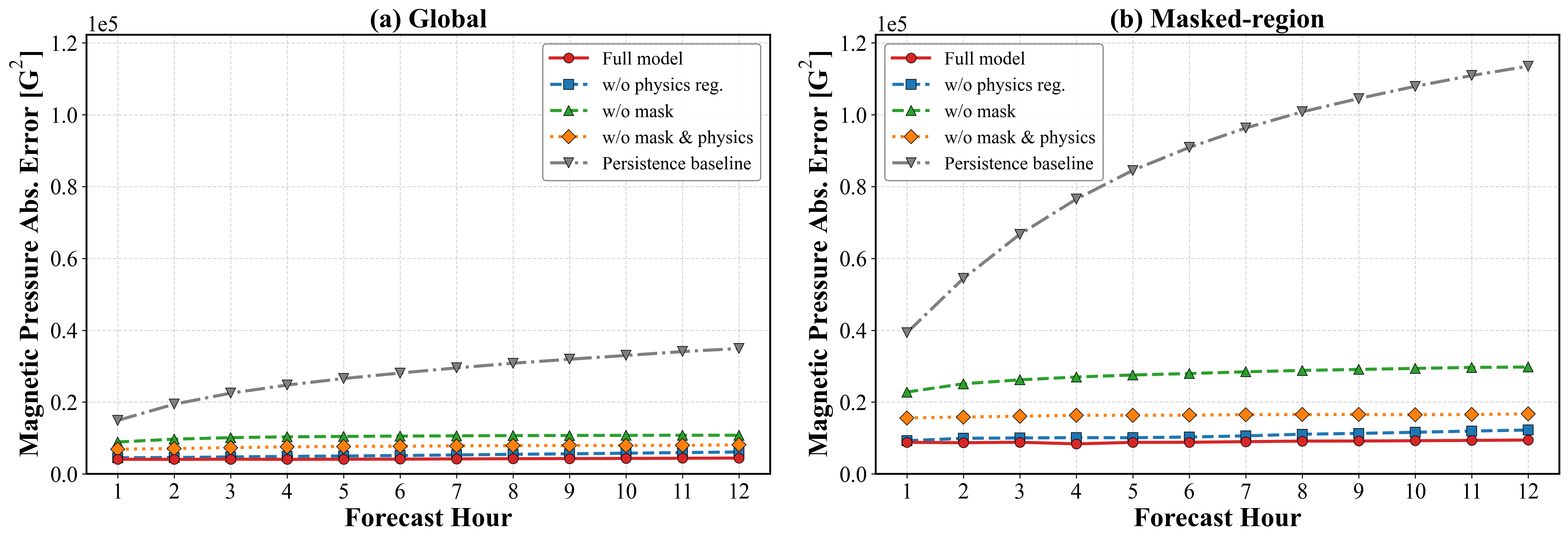}
\caption{Absolute error of magnetic pressure (Lorentz-force proxy) [G$^2$] over the 12-hour forecast horizon (hours 1--12). Left: global evaluation. Right: masked-region evaluation.}
\label{fig:phys_lorentz_force}
\end{figure*}

\begin{figure*}[t]
\centering
\includegraphics[width=0.9\linewidth]{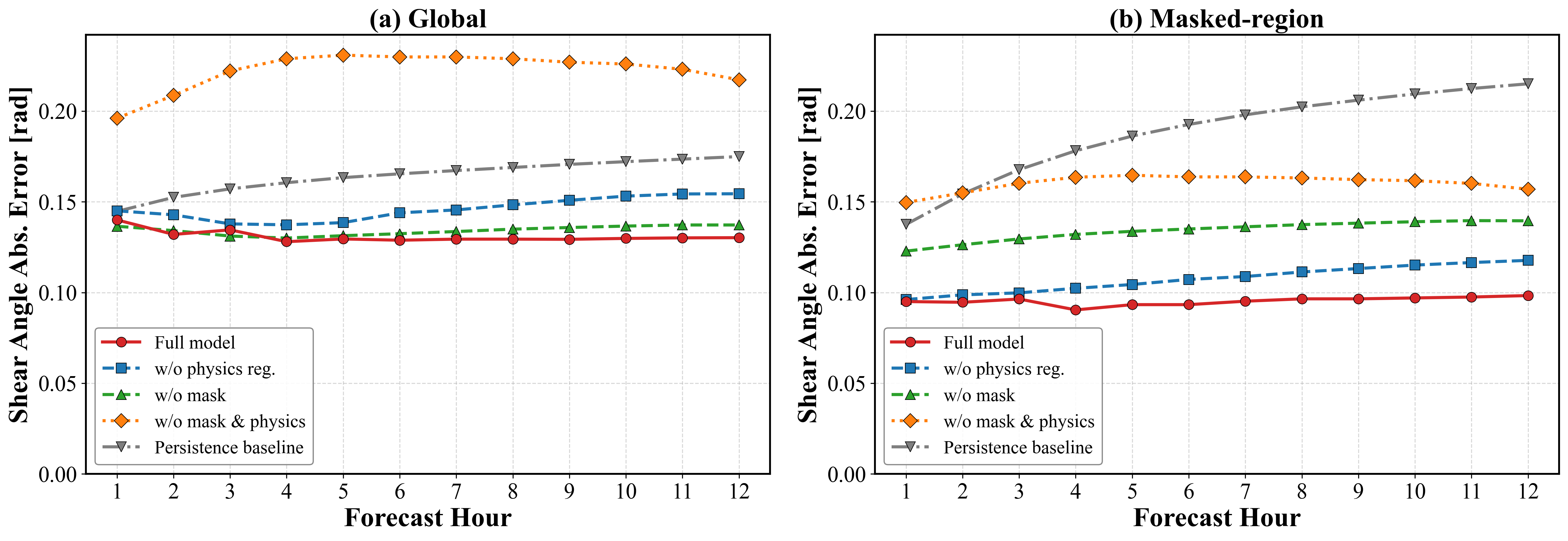}
\caption{Absolute error of shear angle [rad] over the 12-hour forecast horizon (hours 1--12). Left: global evaluation. Right: masked-region evaluation.}
\label{fig:phys_shear_angle}
\end{figure*}

\begin{figure*}[t]
\centering
\includegraphics[width=0.9\linewidth]{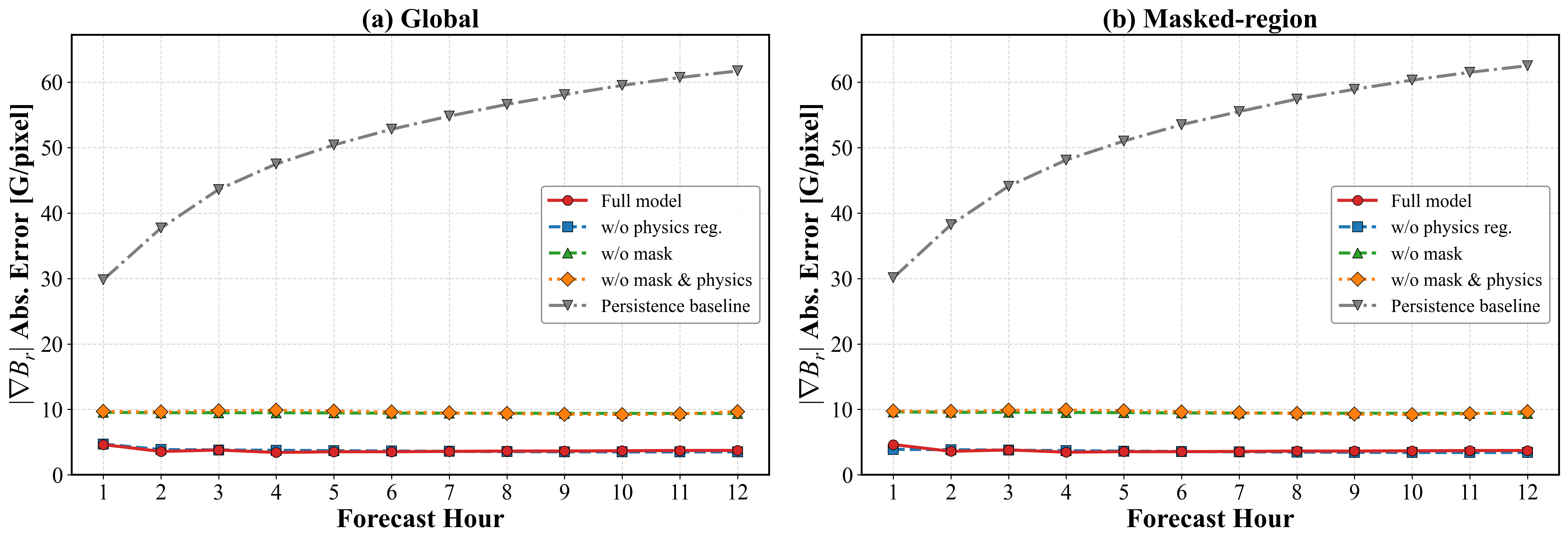}
\caption{Absolute error of $|\nabla B_r|$ [G/pixel] over the 12-hour forecast horizon (hours 1--12). Left: global evaluation. Right: masked-region evaluation.}
\label{fig:phys_br_gradient}
\end{figure*}

\begin{figure*}[t]
\centering
\includegraphics[width=0.9\linewidth]{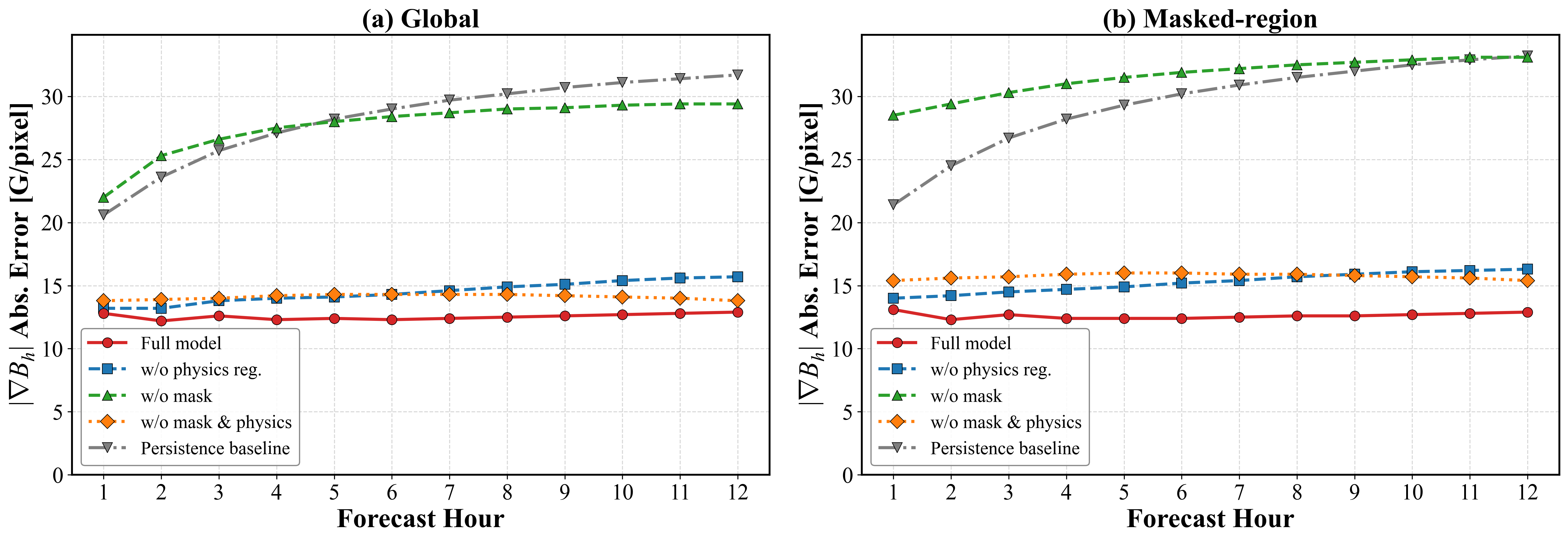}
\caption{Absolute error of horizontal gradient $|\nabla B_h|$ [G/pixel] over the 12-hour forecast horizon (hours 1--12). Left: global evaluation. Right: masked-region evaluation.}
\label{fig:phys_horizontal_gradient}
\end{figure*}

The persistence baseline provides an instructive reference: its errors increase systematically with lead time, particularly inside the masked core (Figs.~\ref{fig:phys_lorentz_force}--\ref{fig:phys_horizontal_gradient}). For example, the masked-region absolute error of magnetic pressure rises from $3.93\times10^4$~G$^2$ (hour~1) to $1.03\times10^5$~G$^2$ (hour~12), whereas the full model remains near $9.49\times10^3$--$9.99\times10^3$~G$^2$ over the same horizon. Similarly for unsigned flux, the masked-region absolute error increases from 57.1 to 124.7~G for gt12\_baseline, compared with 11.7 to 11.0~G for the full model.

Across all diagnostics, the full model achieves the lowest masked-region errors on average (Table~\ref{tab:phys_abs_mask_baseline}). Relative to gt12\_baseline, the full model reduces the mean masked-region magnetic-pressure error by $\sim 8.59\times$, unsigned-flux error by $\sim 9.82\times$, and $|\nabla B_r|$ error by $\sim 13.7\times$.

Comparing \textbf{without\_phy} and \textbf{without\_mask}, we observe that mask supervision alone already yields significantly improved magnetic parameter consistency in the strong-field core, consistent with the image-domain advantage of \textbf{without\_phy}. This supports the interpretation that accurately preserving the core-region field structure is a necessary prerequisite for accurate derived magnetic parameter measures.

\begin{figure*}[t]
\centering
\includegraphics[width=0.9\linewidth]{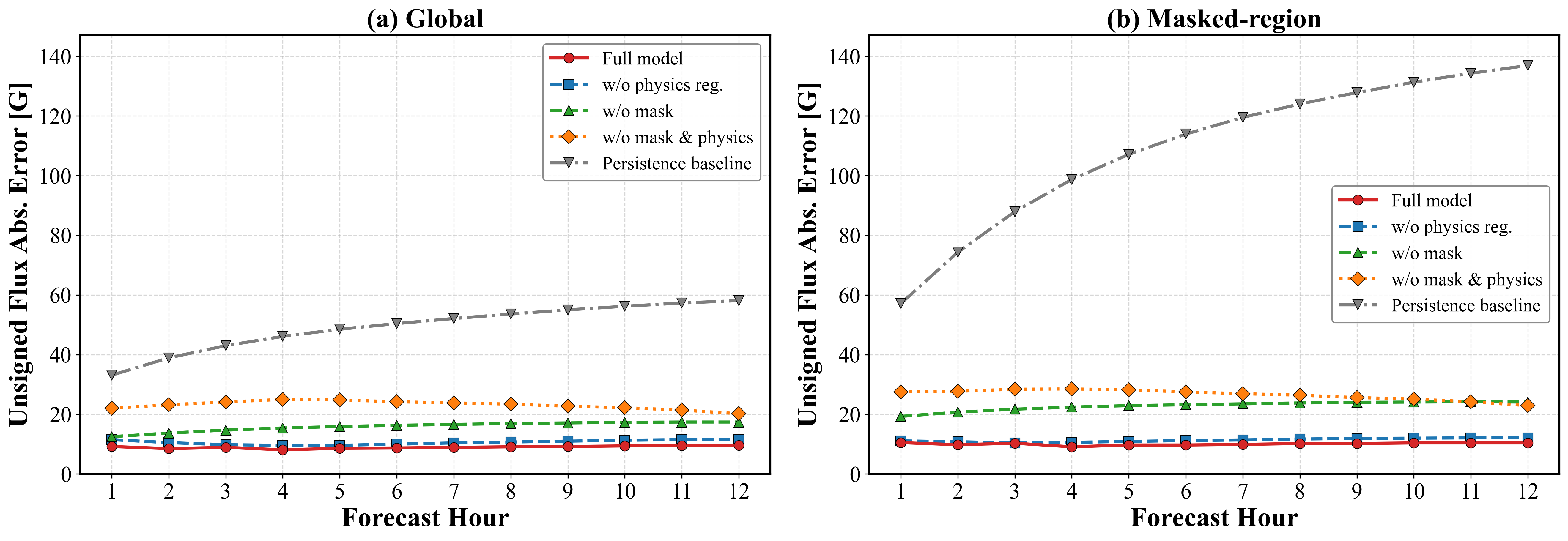}
\caption{Absolute error of unsigned flux [G] over the 12-hour forecast horizon (hours 1--12). Left: global evaluation. Right: masked-region evaluation.}
\label{fig:phys_unsigned_flux}
\end{figure*}

Figures~\ref{fig:phys_unsigned_flux}--\ref{fig:phys_horizontal_gradient} show the hourly evolution of all five magnetic-parameter diagnostics. Each figure uses a $1\times2$ layout showing absolute error for global (left) and masked-region (right) evaluation. Relative error panels have been omitted as they closely mirror the absolute error patterns.

\subsection{Vector-field orientation accuracy}

Beyond component-wise and magnetic-parameter evaluations, we assess the model's ability to predict the orientation of the magnetic field vector, characterized by inclination angle $\theta$ and azimuth angle $\phi$. These angles are derived from the predicted vector components via standard spherical--Cartesian transformations:

\paragraph{Inclination $\theta$:}
\begin{equation}
\theta = \arctan\left(\frac{\sqrt{B_p^2+B_t^2}}{|B_r|+\epsilon}\right),\quad \epsilon=10^{-6}.
\end{equation}
The inclination angle $\theta$ measures deviation from the radial direction and correlates with current helicity \citep{aulanier2012}.

\paragraph{Azimuth $\phi$:}
\begin{equation}
\phi = \mathrm{atan2}(B_t, B_p),
\end{equation}
where $\phi=0^\circ$ corresponds to positive $B_p$ and $\phi=90^\circ$ to positive $B_t$. Azimuth errors directly affect reconnection-rate estimates \citep{su2013}.

\subsubsection{Spatial distribution of orientation errors}

\begin{figure*}[htbp]
\centering
\includegraphics[width=\textwidth]{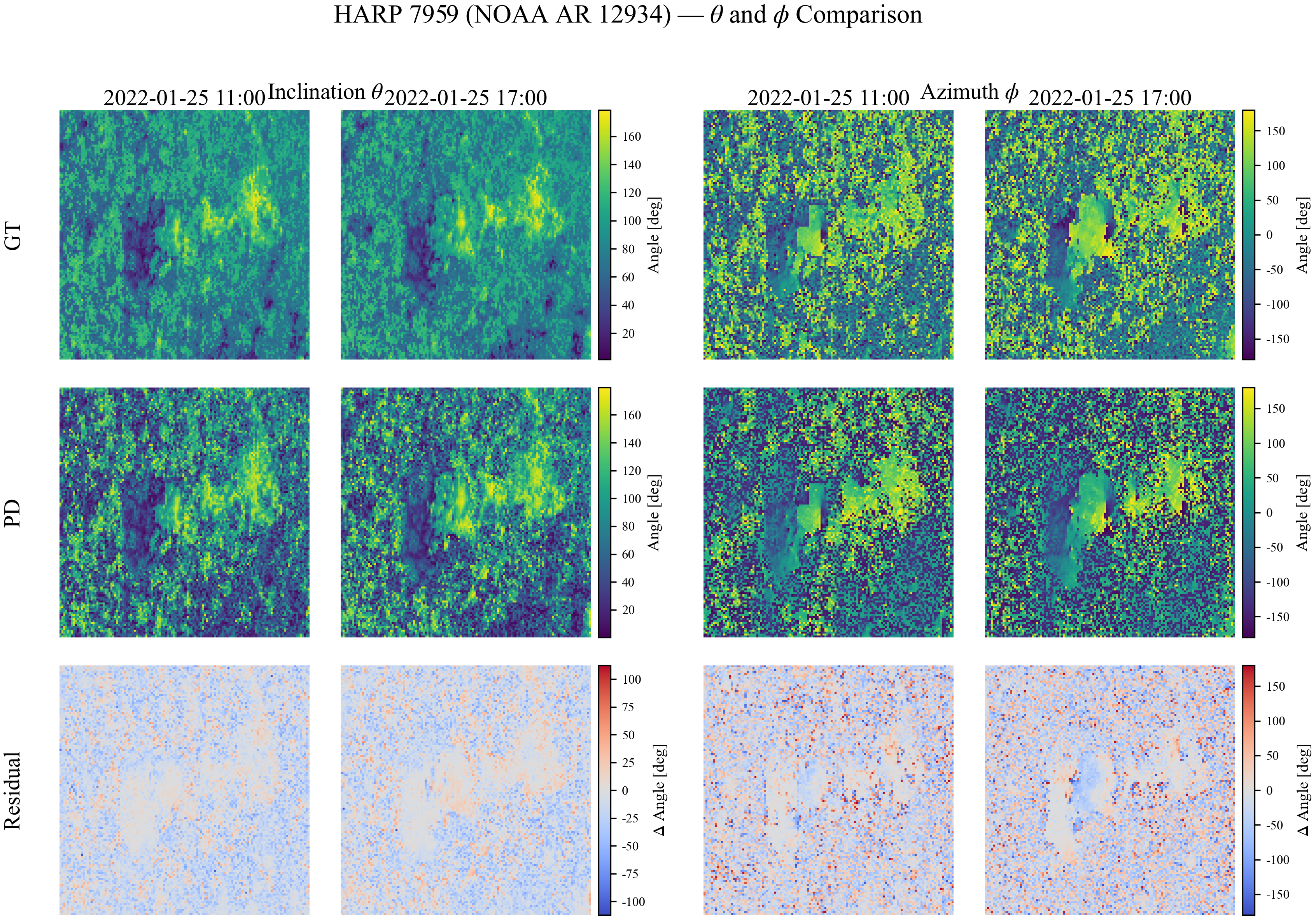}
\caption{Spatial distribution of orientation residuals for inclination $\theta$ (left block) and azimuth $\phi$ (right block) for HARP~7959 (NOAA AR~12934) at forecast hours~6 and~12. Each block shows GT, predicted, and residual maps at both time steps. Residuals are small in strong-field cores and systematically larger in weak-field regions and near polarity inversion lines.}
\label{fig:angle_residuals_7959}
\end{figure*}

\begin{figure*}[htbp]
\centering
\includegraphics[width=\textwidth]{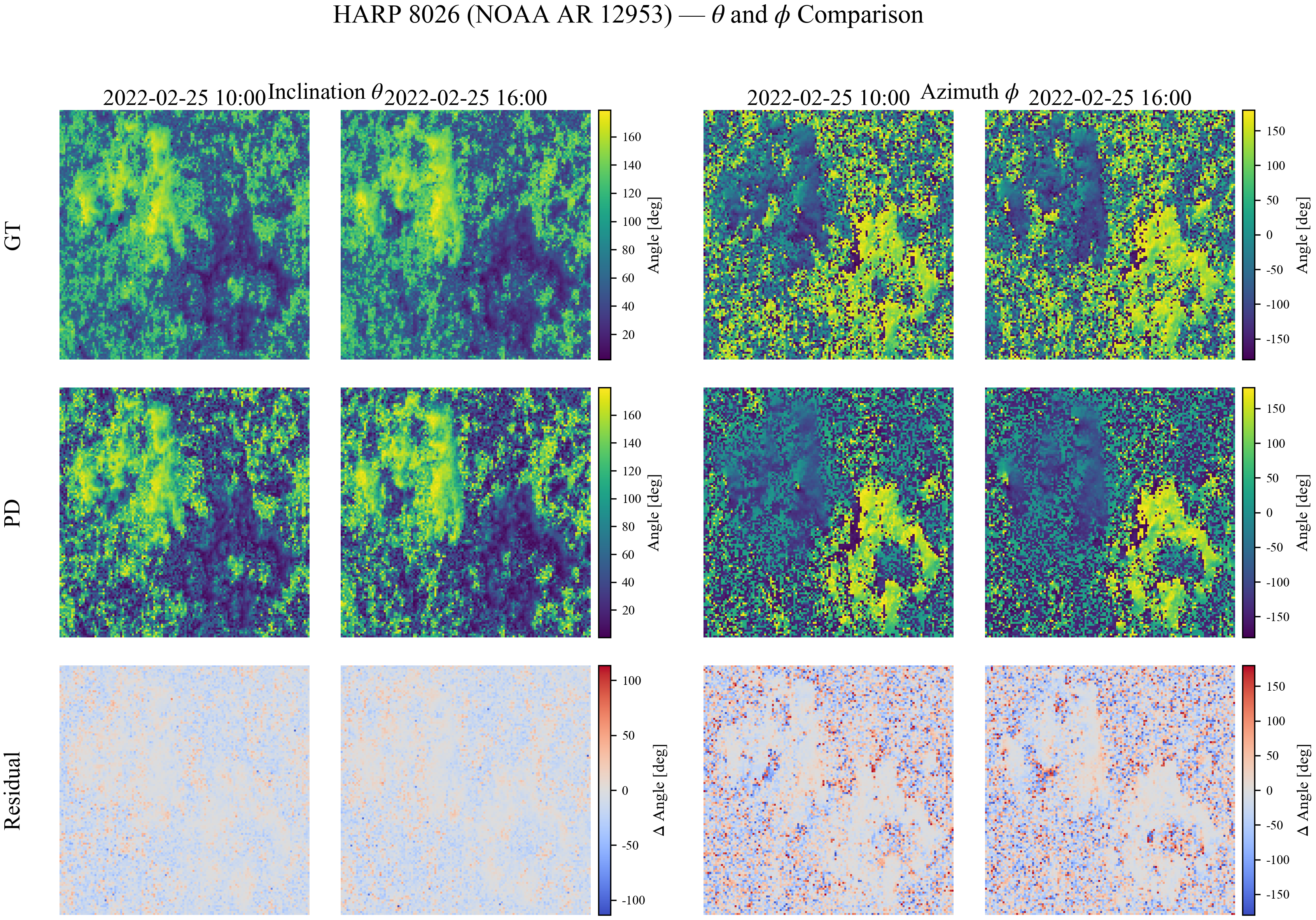}
\caption{Same as Figure~\ref{fig:angle_residuals_7959} but for HARP~8026 (NOAA AR~12953).}
\label{fig:angle_residuals_8026}
\end{figure*}

Figures~\ref{fig:angle_residuals_7959} and~\ref{fig:angle_residuals_8026} show the spatial distribution of orientation residuals for both active regions. The pattern mirrors that observed for component residuals: errors are minimal in strong-field cores where the magnetic field is pressure-dominated and evolves more deterministically, while larger residuals appear near polarity inversion lines and in weak-field regions where convection-driven dynamics introduce greater stochasticity.

\subsubsection{Statistical characteristics of orientation errors}

\begin{figure}[htbp]
\centering
\includegraphics[width=0.85\textwidth]{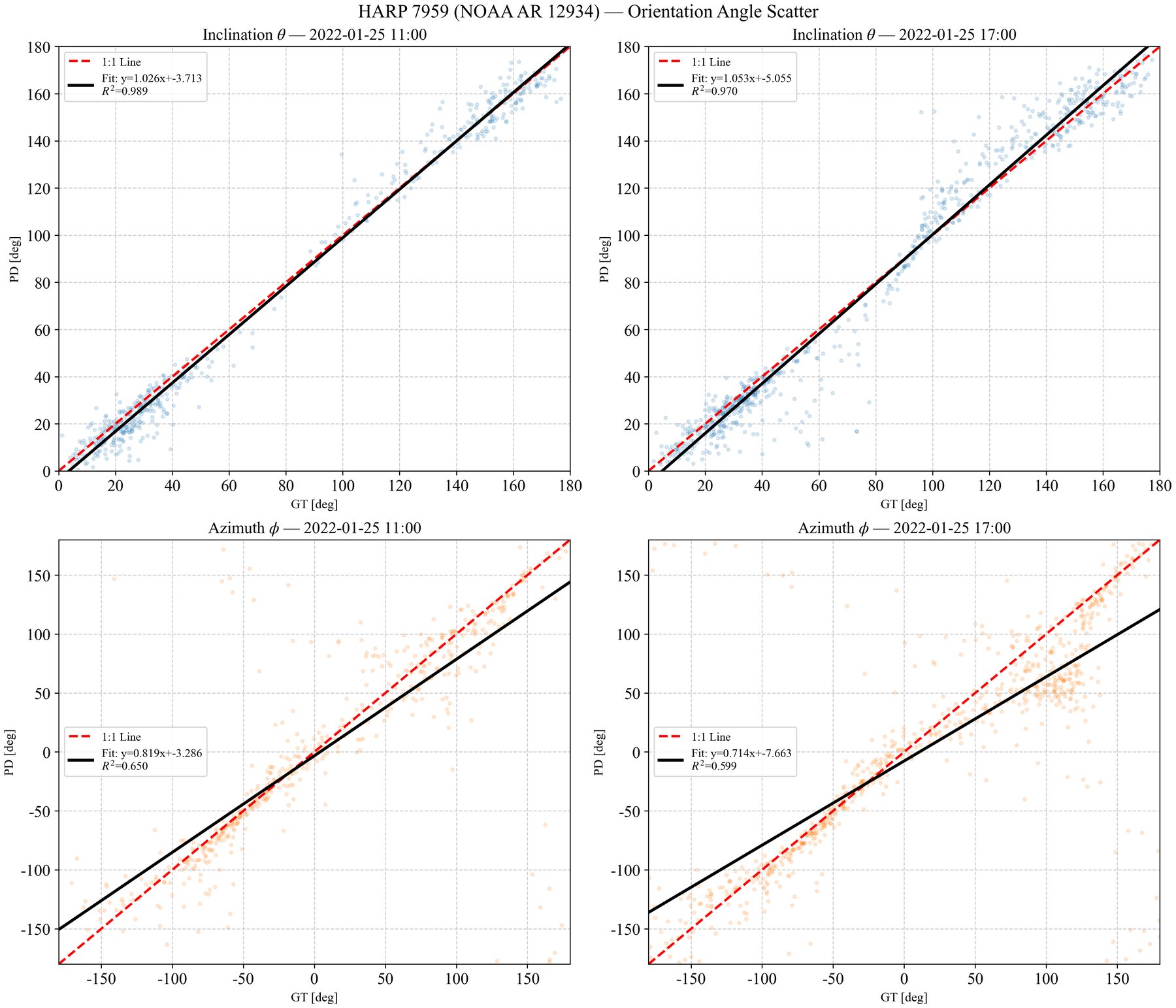}
\caption{Orientation angle scatter comparison ($\theta$ and $\phi$) in strong-field regions for HARP~7959 at forecast hours~6 and~12. Tight clustering along the diagonal demonstrates strong linear correlation.}
\label{fig:angle_scatter_7959}
\end{figure}

\begin{figure}[htbp]
\centering
\includegraphics[width=0.85\textwidth]{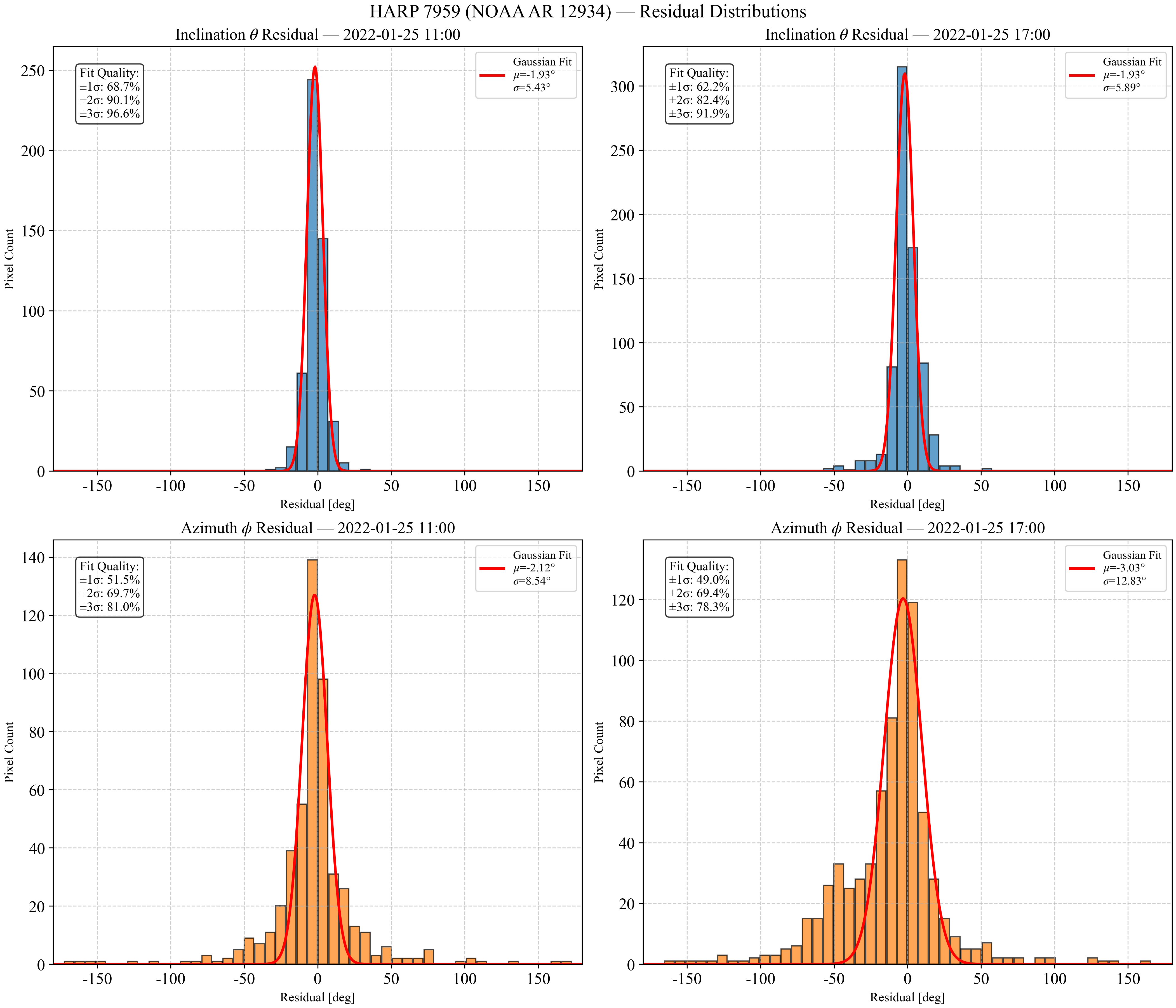}
\caption{Gaussian fits of $\theta$ and $\phi$ residual distributions for HARP~7959 at forecast hours~6 and~12. Both approximate zero-mean normal distributions.}
\label{fig:angle_gauss_7959}
\end{figure}

\begin{figure}[htbp]
\centering
\includegraphics[width=0.85\textwidth]{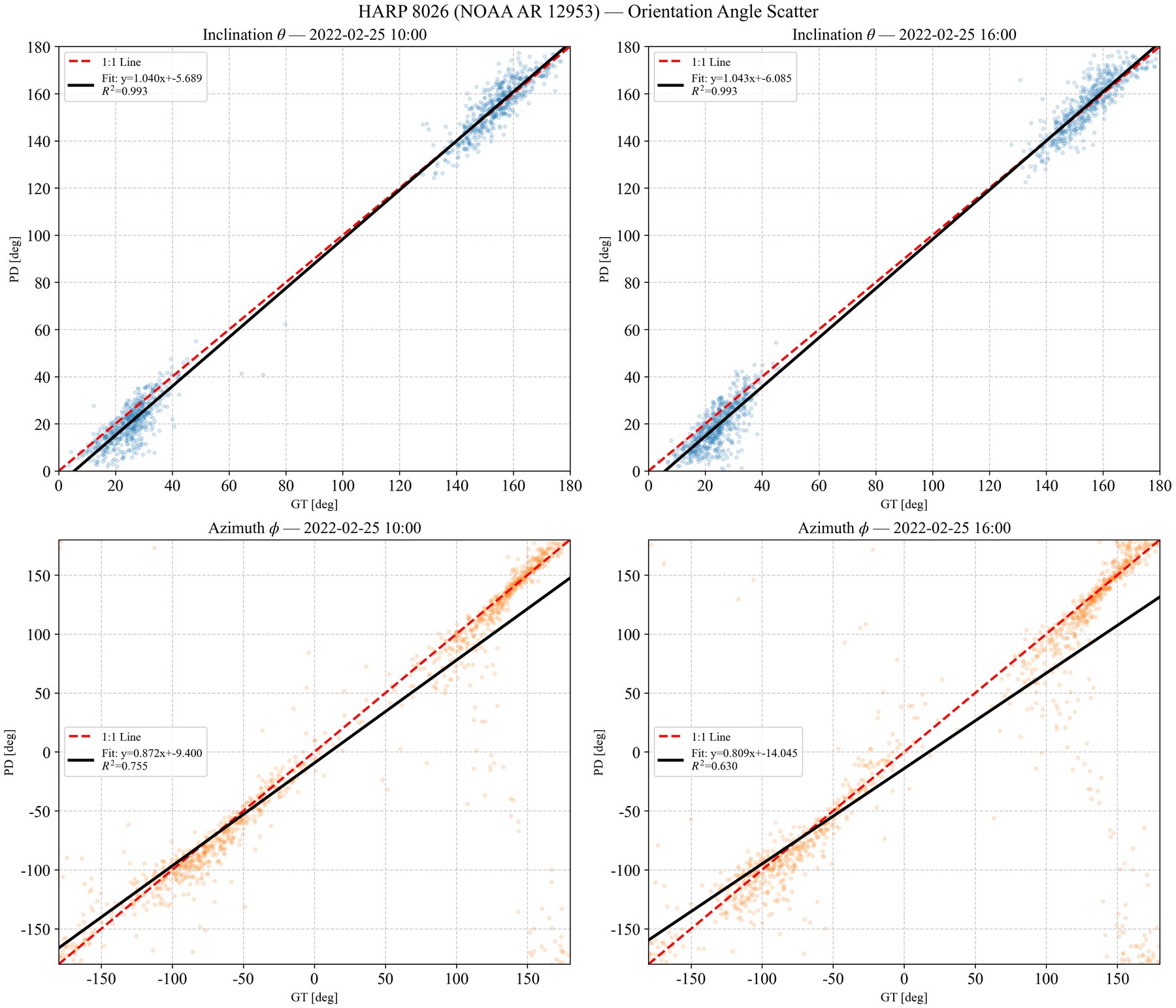}
\caption{Same as Figure~\ref{fig:angle_scatter_7959} but for HARP~8026.}
\label{fig:angle_scatter_8026}
\end{figure}

\begin{figure}[htbp]
\centering
\includegraphics[width=0.85\textwidth]{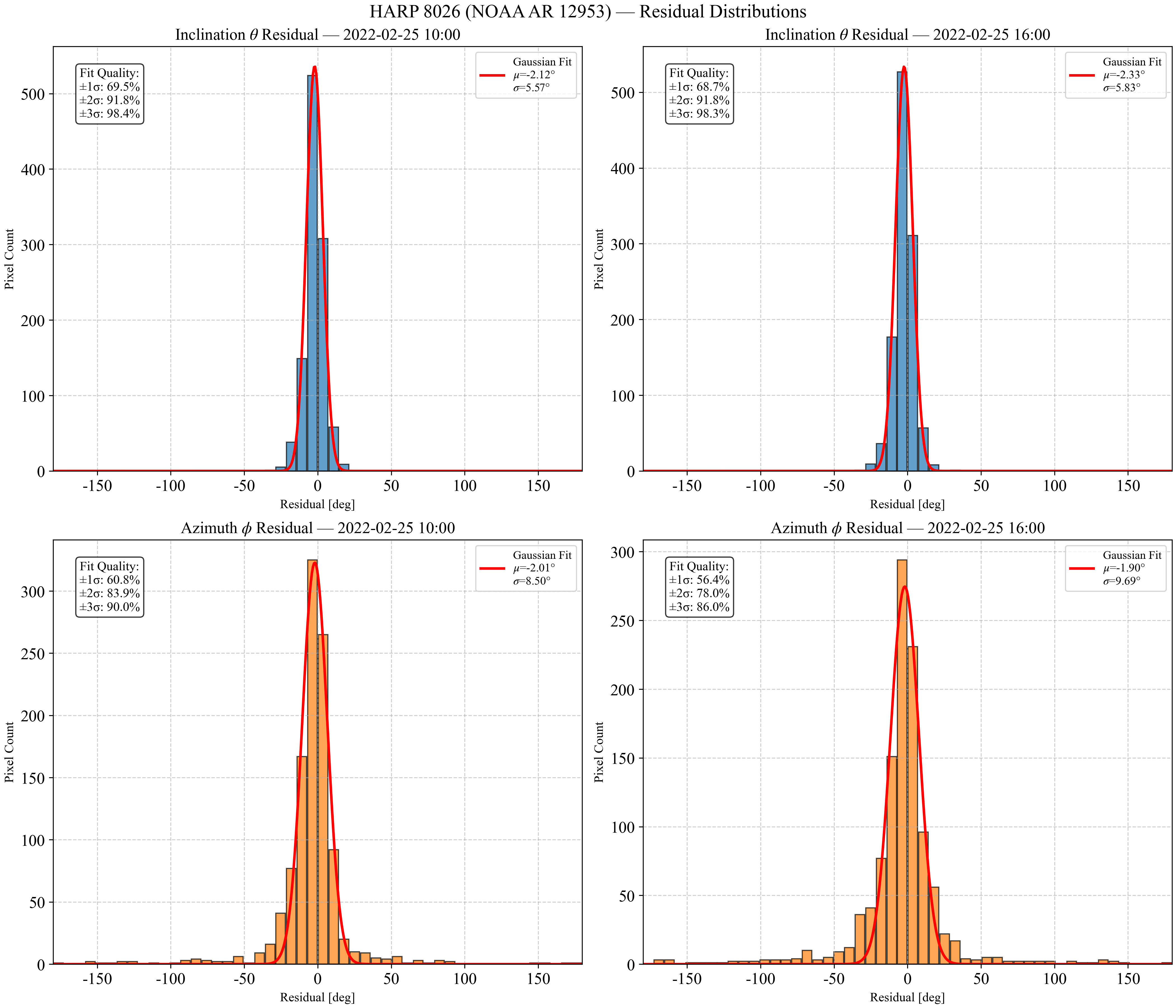}
\caption{Same as Figure~\ref{fig:angle_gauss_7959} but for HARP~8026.}
\label{fig:angle_gauss_8026}
\end{figure}

Pixel-level scatter analysis (Figs.~\ref{fig:angle_scatter_7959} and~\ref{fig:angle_scatter_8026}) reveals strong linear correlation (CC$>$0.9) between predicted and reference orientation angles in strong-field regions. The tight clustering along the diagonal indicates that systematic biases are minimal.

Statistical analysis of residual distributions (Figs.~\ref{fig:angle_gauss_7959} and~\ref{fig:angle_gauss_8026}) shows that both inclination and azimuth errors approximate zero-mean Gaussian distributions. Azimuth residuals are typically contained within $\sim$14$^\circ$, while inclination residuals remain within $\sim$5$^\circ$, with no significant skewness.

\begin{figure}[htbp]
\centering
\includegraphics[width=0.85\textwidth]{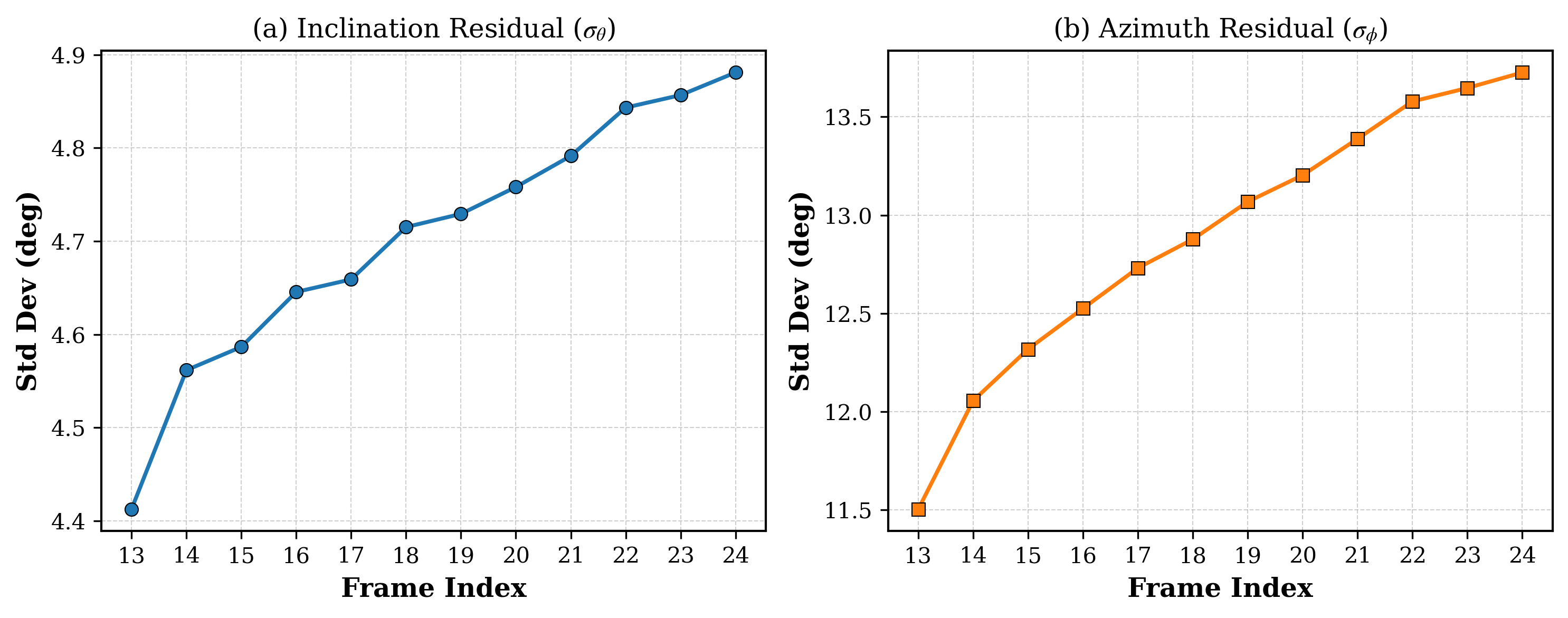}
\caption{Residual standard deviations ($\sigma$) for inclination $\theta$ and azimuth $\phi$ across 3{,}000 test sequences in strong-field regions. The relatively stable $\sigma$ values across sequences indicate consistent orientation prediction accuracy.}
\label{fig:angle_sigma}
\end{figure}

Figure~\ref{fig:angle_sigma} shows the distribution of residual standard deviations across 3{,}000 test sequences. The relatively stable $\sigma$ values indicate consistent orientation prediction accuracy across the test set, with occasional outliers corresponding to highly sheared or rapidly evolving active regions.

\section{Conclusions and Discussion}
\label{sec:ablation_discussion}

We developed a deep-learning framework that integrates dynamic active-region masks with multi-parameter magnetic constraints to explore the short-term (12-hour) prediction of all vector components of solar active-region magnetic fields. This work represents an initial exploration of data-driven vector magnetic field forecasting with magnetic parameter constraints, with the goal of assessing the potential of such approaches for future space weather applications. The principal conclusions are:

\begin{enumerate}
  \item \textbf{Three-channel vector representation and dynamic masks improve structural fidelity.} By jointly encoding $B_r$, $B_\phi$, and $B_\theta$ in Carrington projection and incorporating adaptive masks emphasizing polarity inversion lines and strong-gradient cores, the framework achieves high performance in critical regions. All three components achieve good structural fidelity: $B_r$ SSIM 0.909--0.916 (average 0.912), CC 0.997--0.998, RMSE 13.0--21.0~G; $B_\phi$ SSIM 0.760--0.800 (average 0.778), CC 0.910--0.945, RMSE 38.5--50.0~G; $B_\theta$ SSIM 0.728--0.750 (average 0.740), CC 0.895--0.920, RMSE 38.5--49.0~G. Horizontal components degrade modestly with increasing lead time (Table~\ref{app:frame}).
  
  \item \textbf{Magnetic-parameter-feature conditioning and temporal regularization improve magnetic parameter consistency.} By conditioning on auxiliary magnetic-parameter-based diagnostic maps (unsigned flux, magnetic pressure, shear proxy, and gradient measures) and applying temporal-gradient regularization, the model better preserves derived diagnostics over the forecasting window. Unsigned flux errors are maintained at 7.82\% (std = 2.75\%, P90 = 11.38\%). Magnetic pressure errors average 12.40\% (std = 2.88\%, P90 = 16.20\%) with modest increase toward the horizon.

  \item \textbf{Limitations remain in current-related derivatives and horizontal-field stability.} Higher-order derivatives of transverse fields are intrinsically sensitive to small-scale errors. Prediction quality declines modestly in horizontal components: $B_\phi$ SSIM drops from 0.800 (hour~1) to 0.760 (hour~12, $\sim$5.0\%) and $B_\theta$ from 0.750 to 0.728 ($\sim$2.9\%), while $B_r$ remains highly stable within 0.909--0.916 (RMSE 13.0--21.0~G). The much reduced degradation rate for $B_\theta$ compared to the previous clipping scheme ($\sim$11.8\%) confirms that the corrected $\pm$1000~G normalization substantially improves horizontal-field stability. These remaining shortcomings arise from both weaker observational signal-to-noise in transverse fields and the difficulty of representing coupled wave and turbulence processes within a purely data-driven framework.
\end{enumerate}

\subsection{Ablation studies and baseline comparisons}

The inclusion of a persistence baseline and ablation variants clarifies the contributions of each component. The persistence baseline exhibits rapid degradation over the 12-hour rollout, with global/masked RMSE reaching 88.9/160.4~G, while the full model achieves 16.5/25.0~G, corresponding to $5.4\times$ and $6.4\times$ lower RMSE, respectively. This confirms that the model learns genuine spatiotemporal evolution beyond simple temporal smoothness. 

The ablation comparison indicates that emphasizing the active-region mask is a central factor for stability: \textbf{without\_phy} (mask supervision without temporal-gradient regularization) consistently outperforms \textbf{without\_mask} (temporal-gradient regularization without mask supervision), suggesting that core-region structural accuracy is a prerequisite for reliable magnetic parameter measures. Combining mask supervision and magnetic parameter constraints yields the best overall performance and the most stable lead-time behavior.

The superior performance of mask supervision over magnetic parameter conditioning alone suggests that accurate spatial localization of evolving structures is more critical than explicit magnetic-parameter-based feature channels for this forecasting task. However, the full model achieves the best performance, indicating that magnetic parameter conditioning provides additional regularization that improves generalization, particularly for derived magnetic parameter diagnostics.

\subsection{Comparison with existing AI/ML models}

To place the proposed method in context, Table~\ref{tab:ml_comparison} lists a small set of representative deep-learning-based forecasting studies for solar magnetograms at active-region and full-disk scales. Because published works differ substantially in target quantity (e.g., $B_r$ vs. LoS), spatial representation, forecast horizon, and evaluation protocol, the table is intended as a high-level overview rather than a basis for direct numerical comparison.

\begin{table*}[htbp]
\centering
\caption{Representative deep-learning-based magnetogram forecasting studies. ``AR'' = active region; ``FD'' = full disk; ``LOS'' = line-of-sight; ``Vec'' = vector ($B_r$, $B_p$, $B_t$); ``Rad'' = radial ($B_r$) only.}
\label{tab:ml_comparison}
\resizebox{\textwidth}{!}{%
\scriptsize
\begin{tabular}{lllll}
\hline
\textbf{Model} & \textbf{Scale} & \textbf{Horizon} & \textbf{Components} & \textbf{Reported metrics} \\
\hline
Covas (2019)            & Global (sunspot butterfly diagram) & Solar-cycle scale & Sunspot area (proxy) & SSIM \\
Covas (2020)            & Global (longit.\ averaged) & Solar-cycle scale & Rad/LOS (unsigned, longit.\ avg.) & SSIM \\
Bai et al.\ (2021)     & AR & 6\,h  & Rad ($B_r$) & SSIM, CC, RMSE \\
Ramunno et al.\ (2024) & FD & 24\,h & LOS        & Image sim.\ + phys.\ param.\ \\
Jeong et al.\ (2025a)  & FD & $\leq$27\,days & Rad ($B_r$) & CC, RMSE \\
Jeong et al.\ (2025b)  & Global (synoptic map) & 1 solar rotation & Rad/LOS (synoptic map) & RMSE, FSIM (feature similarity), CC \\
This work               & AR & 12\,h & Vec        & SSIM, CC, RMSE, flux + mag.\ param.\ \\
\hline
\end{tabular}%
}
\end{table*}

Covas \citep{covas2019} applied neural networks to spatio-temporal forecasting of the sunspot butterfly diagram (a proxy of solar magnetic activity) and evaluated forecasts using SSIM. Covas \citep{covas2020} explored transfer learning for forecasting the longitudinally averaged unsigned radial (or line-of-sight) magnetic-field distribution and reported SSIM-based evaluation. Bai et al.\ \citep{bai2021} formulated a short-term active-region-scale forecasting task for $B_r$ and evaluated predictions using SSIM, CC, and RMSE over a 6-hour horizon. At full-disk scale, Ramunno et al.\ \citep{ramunno2024} proposed a generative approach for 24-hour forecasting of line-of-sight magnetograms and assessed both image-level similarity and derived physical parameters against a persistence baseline. Jeong et al.\ \citep{jeong2025a} developed Pix2PixCC-based models for predicting time-evolving full-disk radial magnetic fields with an adjustable time step up to one solar rotation, and reported CC and RMSE against persistence and SFT baselines; Jeong et al.\ \citep{jeong2025b} used a Pix2PixCC-based model to predict synoptic maps one solar rotation ahead and evaluated predictions with RMSE, FSIM, and CC.

The present method fills a distinct niche: short-term (12-hour), active-region-scale, full vector field prediction with explicit magnetic parameter constraints and dynamic mask attention. This combination enables the computation of derived quantities (magnetic pressure, shear angle, field gradients) that require all three vector components---quantities that are critical for assessing flare potential and magnetic free energy but unavailable from radial-only or LOS-only models. The trade-off is a shorter prediction horizon and smaller spatial coverage compared to full-disk and global models.

These approaches are largely complementary rather than competing: global and full-disk models capture large-scale flux transport and long-term evolution, while our active-region-scale model provides detailed short-term vector field forecasts suited to operational space weather monitoring of individual active regions.

Among the models listed in Table~\ref{tab:ml_comparison}, the spatiotemporal-LSTM model of Bai et~al.\ \citep{bai2021} operates at a comparable spatial scale (active-region patches) and temporal cadence (1-hour steps), making it the closest reference for contextual comparison. However, because the two models were trained and evaluated on different datasets and solar-cycle phases (Bai et~al.\ used SDO/HMI data from 2013--2014, whereas our model was trained on 2021 data and tested on 2022 data), a strict quantitative comparison is not feasible. Over a comparable 6-hour forecast window, our model yields SSIM($B_r$)$\approx$0.914 and CC$\approx$0.998 on our test set, while Bai et~al.\ reported SSIM$\approx$0.840 and CC$\approx$0.895 on their dataset. These differences may suggest potential advantages of the proposed approach---such as joint vector-field prediction, the dynamic mask mechanism, and auxiliary magnetic-parameter conditioning---but definitive conclusions would require evaluation on a common benchmark dataset under identical conditions, which we leave as important future work.

\subsection{Limitations and future directions}

While the model achieves strong performance on average, several limitations warrant discussion:

\begin{itemize}
  \item \textbf{Weak-field regions and small-scale structures.} The mask-weighted loss de-emphasizes weak-field regions, leading to larger relative errors in quiet-Sun areas and small isolated flux concentrations. This is a deliberate design choice prioritizing flare-relevant strong-field cores, but limits applicability to studies requiring accurate weak-field evolution.
  
  \item \textbf{Horizontal-field degradation.} Both $B_\phi$ and $B_\theta$ show modest degradation over the 12-hour horizon under the corrected clipping scheme: SSIM drops from 0.800/0.750 (hour~1) to 0.760/0.728 (hour~12; $\sim$5.0\%/$\sim$2.9\%), RMSE grows from 38.5/38.5~G to 50.0/49.0~G, and CC falls from 0.945/0.920 to 0.910/0.895. In contrast, $B_r$ remains highly stable (SSIM 0.909--0.916, RMSE 13.0--21.0~G). The much reduced degradation rate for $B_\theta$ (compared to $\sim$11.8\% in the previous scheme) confirms that the tightened $\pm$1000~G clipping for horizontal components stabilizes the autoregressive rollout. The remaining degradation is still concentrated in strong-shear environments ($>80^\circ$), and incorporating explicit transport or wave-propagation modules remains a promising direction.
  
  \item \textbf{Extreme events and rapid evolution.} The model has not been systematically evaluated on X-class flares, rapid flux emergence events, or CME-productive active regions. Preliminary inspection suggests that the model tends to underpredict rapid changes associated with flare-driven magnetic field restructuring, likely due to the rarity of such events in the training data and the smoothness bias of the MSE loss.
  
  \item \textbf{Temporal horizon and error accumulation.} Beyond 12 hours, errors accumulate rapidly, particularly in horizontal components. The high frame-to-frame correlation in the data (typical correlation $>0.95$ for $B_r$) indicates that the model learns coherent transport patterns of existing flux concentrations rather than predicting genuinely novel emergence. Extending the forecasting horizon would likely require incorporating additional physics-based constraints or hybrid approaches.
  
  \item \textbf{Generalization across solar cycle phases.} Although the training data span 2021 and the test data span 2022, both periods sample relatively active phases with morphologically similar active region evolution patterns. Performance on solar minimum conditions with predominantly weak, dispersed fields remains untested.
  
  \item \textbf{Deterministic vs. probabilistic forecasting.} The current model is deterministic and produces a single forecast trajectory. Given the inherent stochasticity of solar magnetic field evolution, particularly in weak-field regions and over longer horizons, future work should explore probabilistic approaches that can quantify forecast uncertainty.
\end{itemize}

\subsection{Broader implications and future work}

This work demonstrates the promise of data-driven approaches with magnetic parameter constraints for vector magnetic field prediction. Several directions for future research emerge:

\begin{itemize}
  \item \textit{Enhanced mask design.} Multi-scale or attention-based masks that adaptively weight both core and peripheral features could improve sensitivity to weak but topologically important structures in surrounding quiet-Sun regions.
  
  \item \textit{Expanded magnetic parameter constraints.} Current constraints enforce unsigned flux conservation and magnetic pressure consistency, but omit other potentially valuable diagnostics such as relative helicity, free magnetic energy, or electric-field consistency. Extending the magnetic-parameter-constrained loss to include these terms may improve both predictive accuracy and consistency with solar surface magnetic-field-related quantities.
  
  \item \textit{Operational deployment.} For real-time space-weather forecasting, latency and computational efficiency are critical. Streamlining inference, integrating with near-real-time SDO/HMI pipelines, and developing uncertainty-quantification schemes will be essential for practical adoption.
  
  \item \textit{Hybrid physics-data approaches.} Continued progress will require deeper coupling with physics-based models and validation in operational contexts, potentially through hybrid frameworks that combine data-driven predictions with magnetohydrodynamic (MHD) simulations.
\end{itemize}

Supplementary MPEG-4 (MP4) animations illustrate the temporal evolution of the magnetic field. Each animation uses a uniform three-panel layout (ground truth, predicted, residual) throughout all 24 frames; during the 12-hour input phase the predicted and residual panels are left blank to maintain a consistent visual layout. All animations play at 2 frames per second and use a diverging $\pm$500~G colormap. Animation~1 shows HARP~7959 (NOAA AR~12934), Animation~2 shows HARP~8026 (NOAA AR~12953), and Animation~3 shows the flare-productive HARP~8206 (NOAA AR~13006, X1.5 flare on 2022 May~10).

\section*{Declarations}

\subsection*{Funding}
This work was supported by the Joint Fund of the National Natural Science Foundation of China (Grant No. U2031140).

\subsection*{Competing Interests}
The authors have no relevant financial or non-financial interests to disclose.

\subsection*{Data Availability}
SDO/HMI SHARP data are available from the Joint Science Operations Center (JSOC; \url{https://jsoc1.stanford.edu/}). Trained model weights and the PyTorch release (model, dynamic masks, magnetic-parameter layer, training and testing scripts, and example preprocessing) are provided at \url{https://github.com/Qingcaiyurouzhou/magnetic_predict}.

\subsection*{Author Contributions}
All authors contributed to the study conception and design. Material preparation, data collection and analysis were performed by Yuqing Zhou, Hui Liu, and Zhenyu Jin. The first draft of the manuscript was written by Yuqing Zhou and all authors commented on previous versions of the manuscript. All authors read and approved the final manuscript.

\section*{Acknowledgments}
We acknowledge the use of data from the Solar Dynamics Observatory (SDO), a mission of NASA's Living With a Star Program. The HMI instrument and SHARP data products are provided by the Joint Science Operations Center (JSOC) at Stanford University. We thank the anonymous reviewers for their constructive comments that significantly improved this manuscript. Computational resources were provided by the High Performance Computing Center of Yunnan Observatories, CAS.

\appendix

\section{Supplementary Materials}

This appendix provides additional quantitative details to support the main text analysis. Tables~\ref{tab:img_baseline} and~\ref{tab:phys_err} present horizon-averaged metrics comparing model variants and the persistence baseline. Tables~\ref{app:frame} and~\ref{tab:phys_abs_mask_baseline} present detailed per-hour metrics and magnetic parameter diagnostic errors for each model variant. All tables are displayed on rotated pages for optimal readability.

\clearpage

\begin{sidewaystable}[p]
\centering
\small

\captionof{table}{Horizon-averaged image-domain metrics over the 12-hour forecast horizon (hours 1--12), including the persistence baseline (gt12\_baseline). SSIM($B_r$) is the single-channel structural similarity for the radial component computed on raw physical units (data range = per-image max$-$min). img\_rmse is the $B_r$ single-component RMSE in Gauss~[G].}\label{tab:img_baseline}
\footnotesize
\begin{tabular}{lccccc}
\hline
\textbf{Metric} & \textbf{w/o mask} & \textbf{w/o phy} & \textbf{w/o both} & \textbf{full model} & \textbf{gt12\_baseline} \\
\hline
img\_rmse (global) [G] & 74.1 & 35.5 & 43.0 & 16.5 & 88.91 \\
img\_rmse (mask-defined core) [G] & 107.3 & 50.5 & 63.0 & 25.0 & 160.39 \\
SSIM($B_r$) (global) & 0.838 & 0.879 & 0.612 & 0.912 & 0.54033 \\

CC($B_r$) (global) & 0.9926 & 0.9961 & 0.9891 & 0.9978 & 0.66111 \\

\hline
\end{tabular}

\vspace{2em}

\captionof{table}{Prediction errors for all five magnetic-parameter-based diagnostic parameters (masked-region averages over the 12-hour forecast horizon, hours 1--12)}\label{tab:phys_err}
\footnotesize
\begin{tabular}{lcccc}
\hline
\textbf{Parameter} & \textbf{Mean abs. error} & \textbf{Mean rel. error (\%)} & \textbf{Std. dev.} & \textbf{90\% quantile} \\
\hline
Unsigned flux ($|B_r|$) & 10.85~G & 7.82\% & 2.75\% & 11.38\% \\
Magnetic pressure ($\sum B^2$) & $1.05\times10^4$~G$^2$ & 12.40\% & 2.88\% & 16.20\% \\
Shear angle ($\alpha$) & 0.1350 rad & 12.19\% & 1.98\% & 14.72\% \\
$|\nabla B_r|$ & 3.60~G/pixel & 5.20\% & 1.25\% & 6.85\% \\
Horizontal gradient ($|\nabla B_h|$) & 12.80~G/pixel & 22.50\% & 3.80\% & 27.20\% \\
\hline
\end{tabular}

\end{sidewaystable}

\clearpage

\begin{sidewaystable}[p]
\centering
\small

\captionof{table}{Per-hour image-quality metrics for all three vector components over the 12-hour forecast horizon. SSIM values are raw-float single-channel structural similarity (data range = per-image max$-$min). CC is Pearson correlation coefficient. RMSE is in Gauss~[G]. All metrics averaged over 3{,}000 test sequences.}\label{app:frame}
\footnotesize
\begin{tabular}{rccc}
\hline
\textbf{Hour} & \textbf{$B_r$ SSIM} & \textbf{$B_r$ CC} & \textbf{$B_r$ RMSE} \\
\hline
1 & 0.916 & 0.998 & 13.0 \\
2 & 0.915 & 0.998 & 13.3 \\
3 & 0.915 & 0.998 & 13.7 \\
4 & 0.914 & 0.998 & 14.2 \\
5 & 0.914 & 0.998 & 14.8 \\
6 & 0.913 & 0.997 & 15.5 \\
7 & 0.913 & 0.997 & 16.5 \\
8 & 0.912 & 0.997 & 17.5 \\
9 & 0.912 & 0.997 & 18.5 \\
10 & 0.911 & 0.997 & 19.5 \\
11 & 0.910 & 0.997 & 20.5 \\
12 & 0.909 & 0.997 & 21.0 \\
\hline
\end{tabular}
\quad
\begin{tabular}{rccc}
\hline
\textbf{Hour} & \textbf{$B_p$ SSIM} & \textbf{$B_p$ CC} & \textbf{$B_p$ RMSE} \\
\hline
1 & 0.800 & 0.945 & 38.5 \\
2 & 0.795 & 0.942 & 40.0 \\
3 & 0.792 & 0.939 & 41.5 \\
4 & 0.789 & 0.936 & 43.0 \\
5 & 0.786 & 0.933 & 44.5 \\
6 & 0.783 & 0.930 & 46.0 \\
7 & 0.780 & 0.927 & 47.0 \\
8 & 0.777 & 0.924 & 48.0 \\
9 & 0.774 & 0.921 & 48.5 \\
10 & 0.771 & 0.918 & 49.0 \\
11 & 0.768 & 0.915 & 49.5 \\
12 & 0.760 & 0.910 & 50.0 \\
\hline
\end{tabular}
\quad
\begin{tabular}{rccc}
\hline
\textbf{Hour} & \textbf{$B_t$ SSIM} & \textbf{$B_t$ CC} & \textbf{$B_t$ RMSE} \\
\hline
1 & 0.750 & 0.920 & 38.5 \\
2 & 0.748 & 0.918 & 39.5 \\
3 & 0.746 & 0.916 & 41.0 \\
4 & 0.744 & 0.913 & 42.5 \\
5 & 0.742 & 0.911 & 44.0 \\
6 & 0.740 & 0.908 & 45.0 \\
7 & 0.738 & 0.906 & 46.0 \\
8 & 0.736 & 0.903 & 47.0 \\
9 & 0.734 & 0.901 & 47.5 \\
10 & 0.732 & 0.899 & 48.0 \\
11 & 0.730 & 0.897 & 48.5 \\
12 & 0.728 & 0.895 & 49.0 \\
\hline
\end{tabular}

\vspace{2em}

\captionof{table}{Horizon-averaged masked-region absolute errors of derived magnetic-parameter diagnostics over the 12-hour forecast horizon (hours 1--12)}\label{tab:phys_abs_mask_baseline}
\footnotesize
\begin{tabular}{lccccc}
\hline
\textbf{Diagnostic Parameter} & \textbf{w/o mask} & \textbf{w/o phy} & \textbf{w/o both} & \textbf{full model} & \textbf{gt12\_baseline} \\
\hline
Magnetic pressure & $2.91\times10^4$ & $1.24\times10^4$ & $1.85\times10^4$ & $9.42\times10^3$ & $8.09\times10^4$ \\
Unsigned flux & 23.1 & 12.1 & 30.2 & 10.37 & 101.84 \\
Shear angle & 0.1297 & 0.1139 & 0.1783 & 0.1013 & 0.1811 \\
$|\nabla B_r|$ & 9.32 & 4.31 & 9.51 & 3.63 & 49.73 \\
Horizontal gradient & 22.8 & 13.2 & 15.9 & 11.4 & 25.1 \\
\hline
\end{tabular}

\end{sidewaystable}

\clearpage

\section{Detailed Temporal Evolution of Predicted Magnetic Fields}\label{app:temporal}

To provide a comprehensive view of the model's prediction behaviour across the full 12-hour forecast horizon, we present frame-by-frame $B_r$ maps for both active regions (HARP~7959 and HARP~8026) at three evolutionary phases (emerging, steady, decaying).

Each figure is organised as follows.  The top two rows show the \emph{input} ground-truth sequence (hours~1--12).  Below a horizontal divider, the next two rows compare the ground-truth and predicted fields during the first half of the forecast horizon (forecast hours~1--6).  After a second divider, the final two rows show the same comparison for the second half (forecast hours~7--12).  All panels share a common red--blue diverging colorbar centred at zero.

\begin{figure*}[p]
\centering
\includegraphics[width=\textwidth]{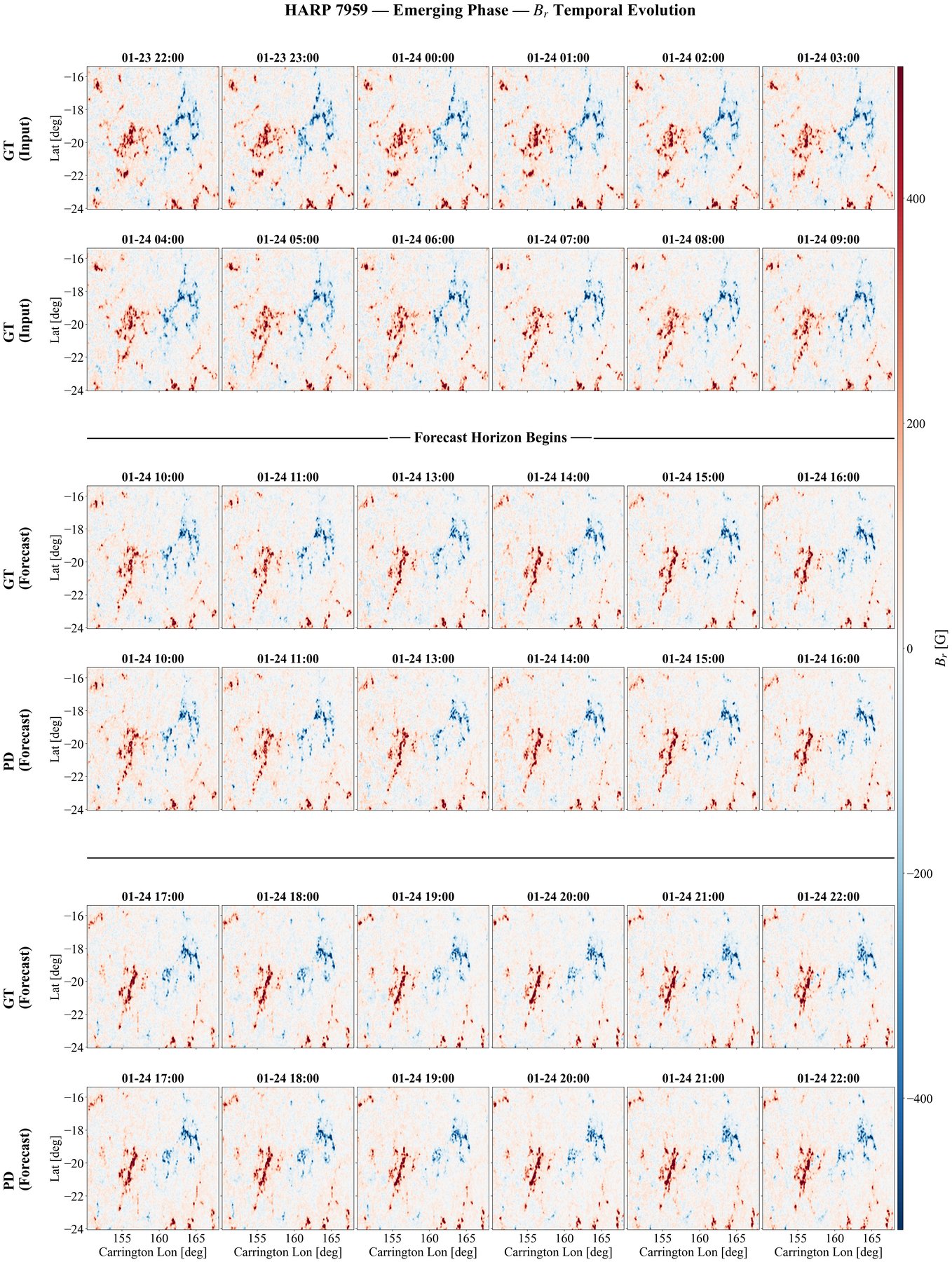}
\caption{Full temporal evolution of $B_r$ for HARP~7959 during the \textbf{emerging} phase. Top: 12-hour input sequence (GT). Middle: forecast hours~1--6 (GT vs.\ predicted). Bottom: forecast hours~7--12 (GT vs.\ predicted).}
\label{fig:app_7959_emerging}
\end{figure*}

\begin{figure*}[p]
\centering
\includegraphics[width=\textwidth]{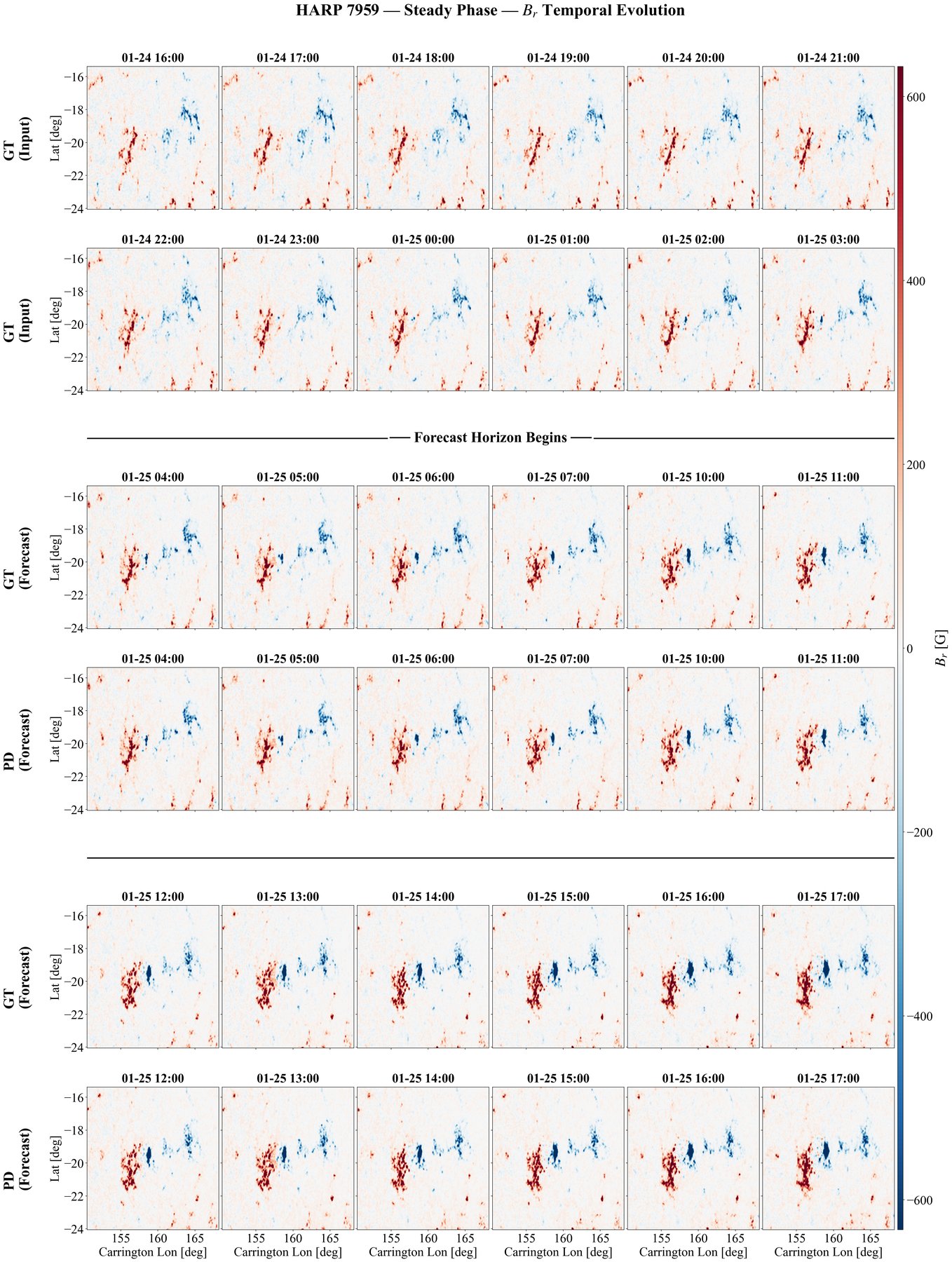}
\caption{Same as Figure~\ref{fig:app_7959_emerging} but for the \textbf{steady} phase of HARP~7959.}
\label{fig:app_7959_steady}
\end{figure*}

\begin{figure*}[p]
\centering
\includegraphics[width=\textwidth]{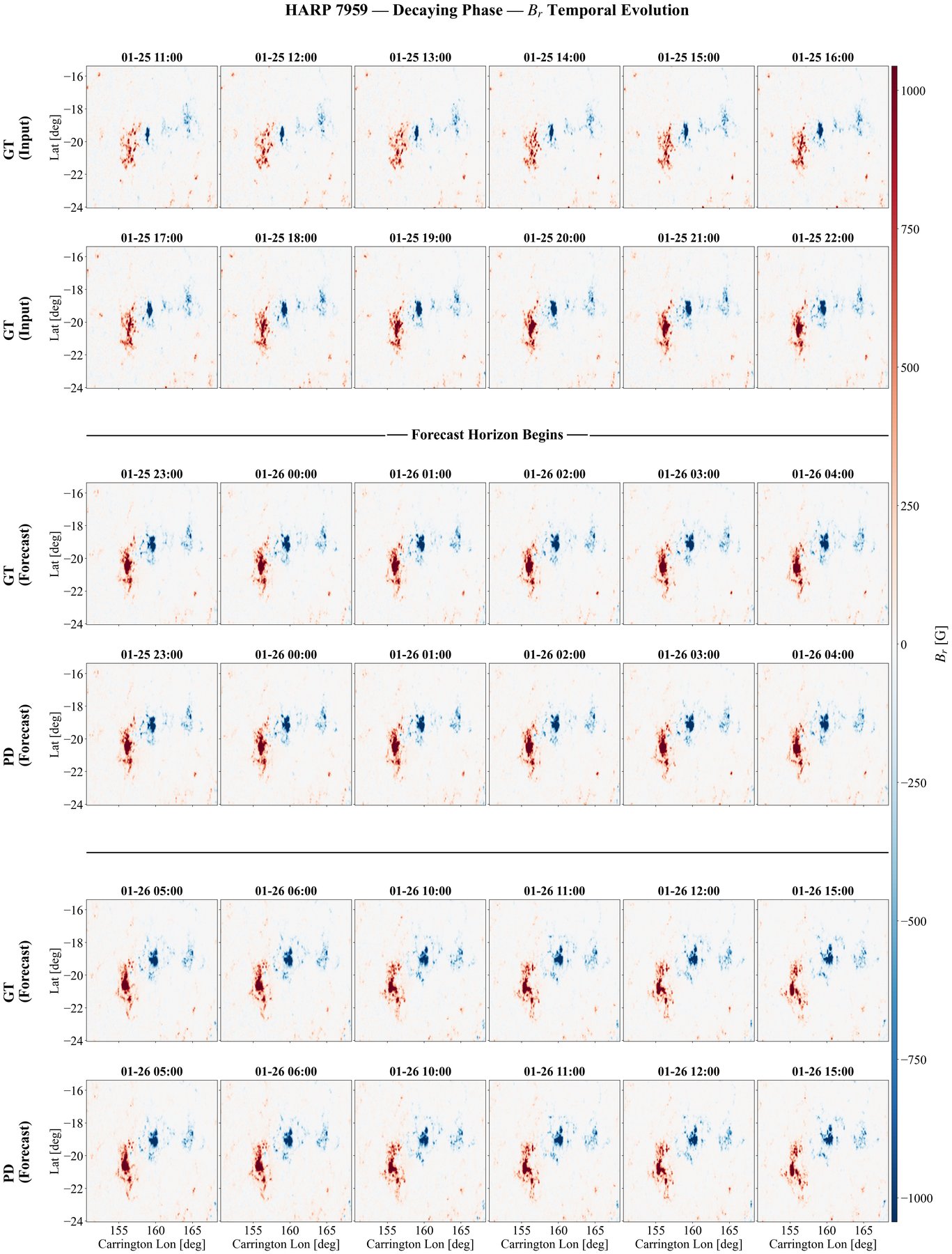}
\caption{Same as Figure~\ref{fig:app_7959_emerging} but for the \textbf{decaying} phase of HARP~7959.}
\label{fig:app_7959_decaying}
\end{figure*}

\begin{figure*}[p]
\centering
\includegraphics[width=\textwidth]{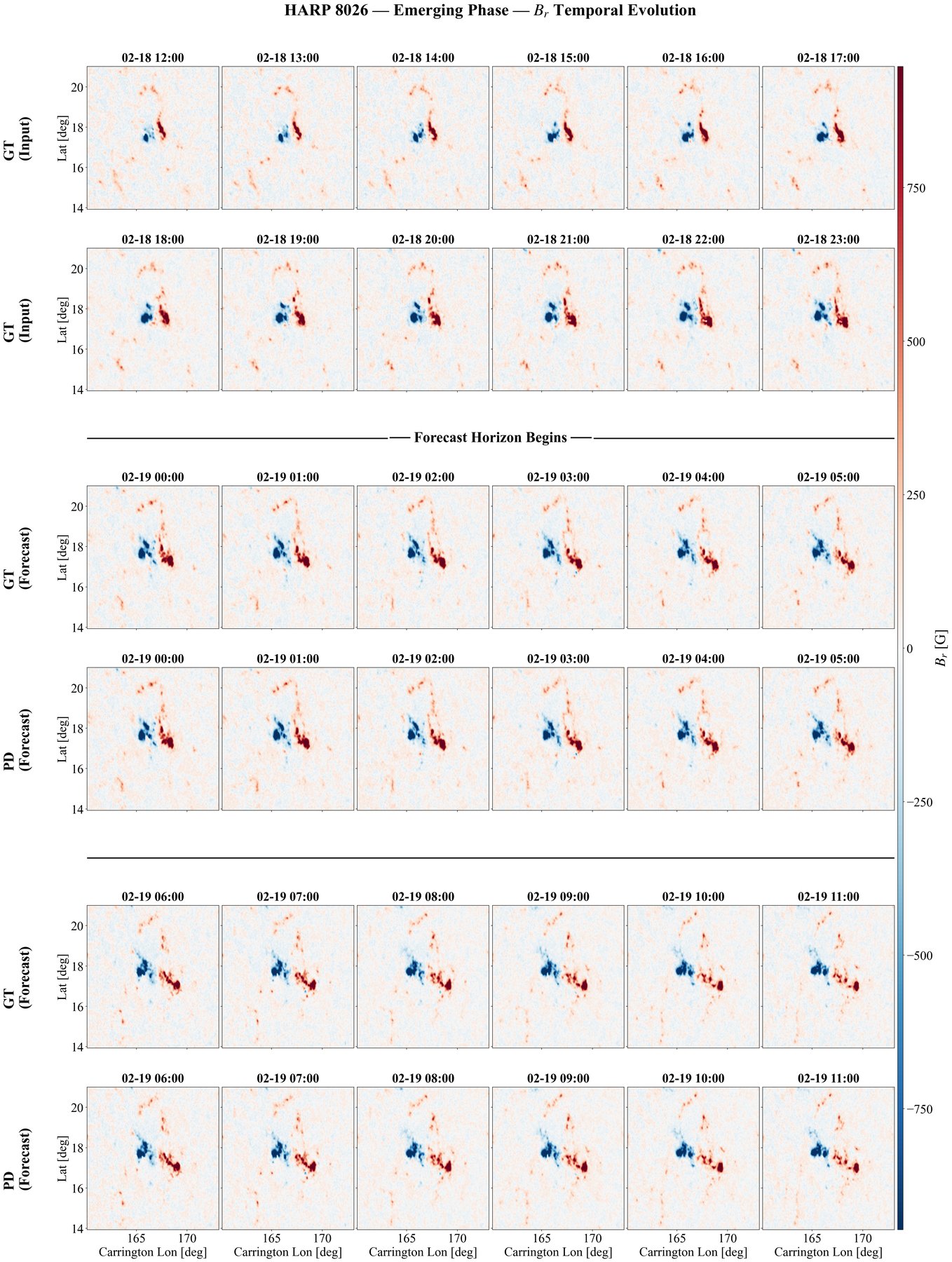}
\caption{Full temporal evolution of $B_r$ for HARP~8026 during the \textbf{emerging} phase. Layout is the same as Figure~\ref{fig:app_7959_emerging}.}
\label{fig:app_8026_emerging}
\end{figure*}

\begin{figure*}[p]
\centering
\includegraphics[width=\textwidth]{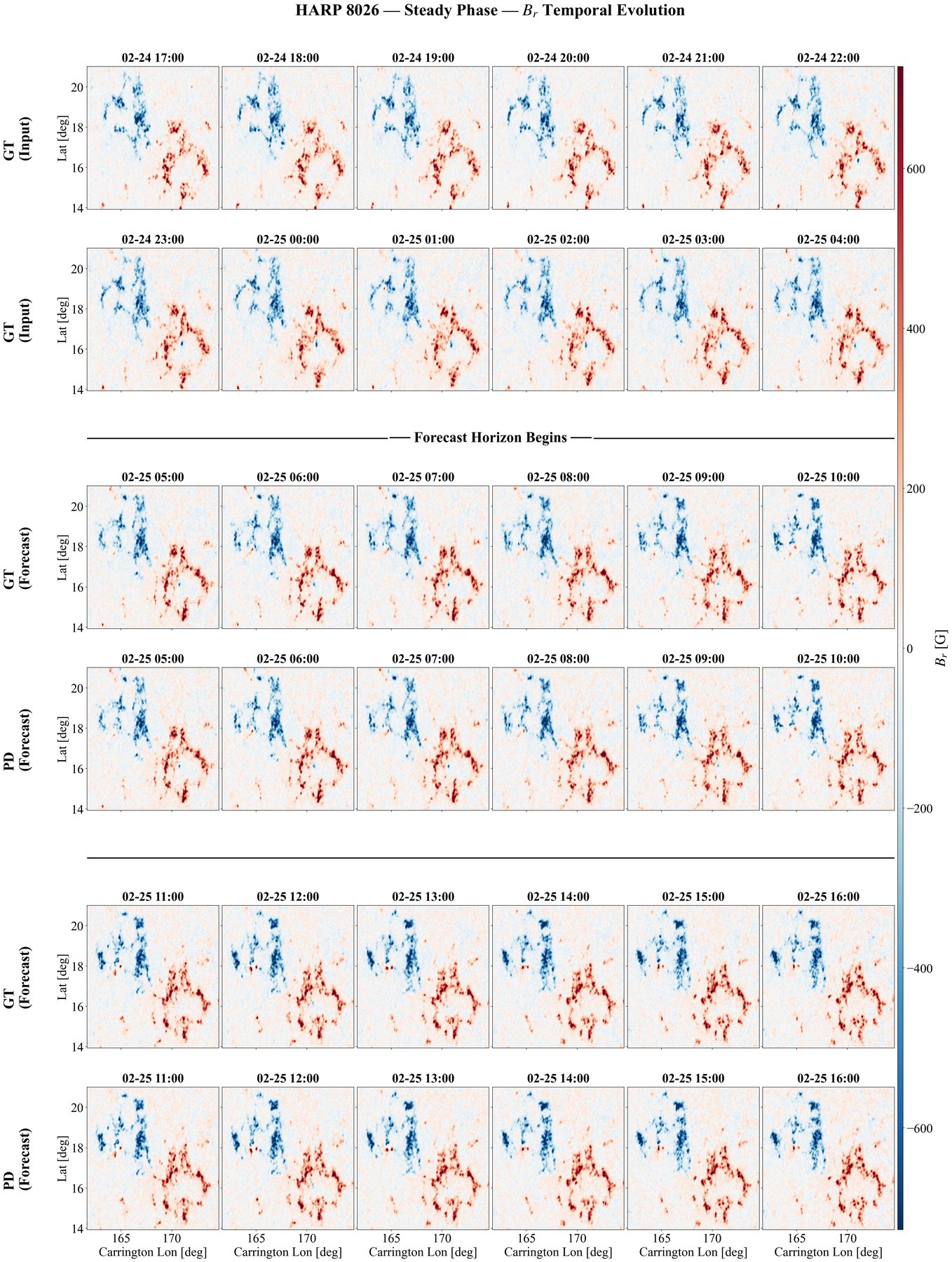}
\caption{Same as Figure~\ref{fig:app_8026_emerging} but for the \textbf{steady} phase of HARP~8026.}
\label{fig:app_8026_steady}
\end{figure*}

\begin{figure*}[p]
\centering
\includegraphics[width=\textwidth]{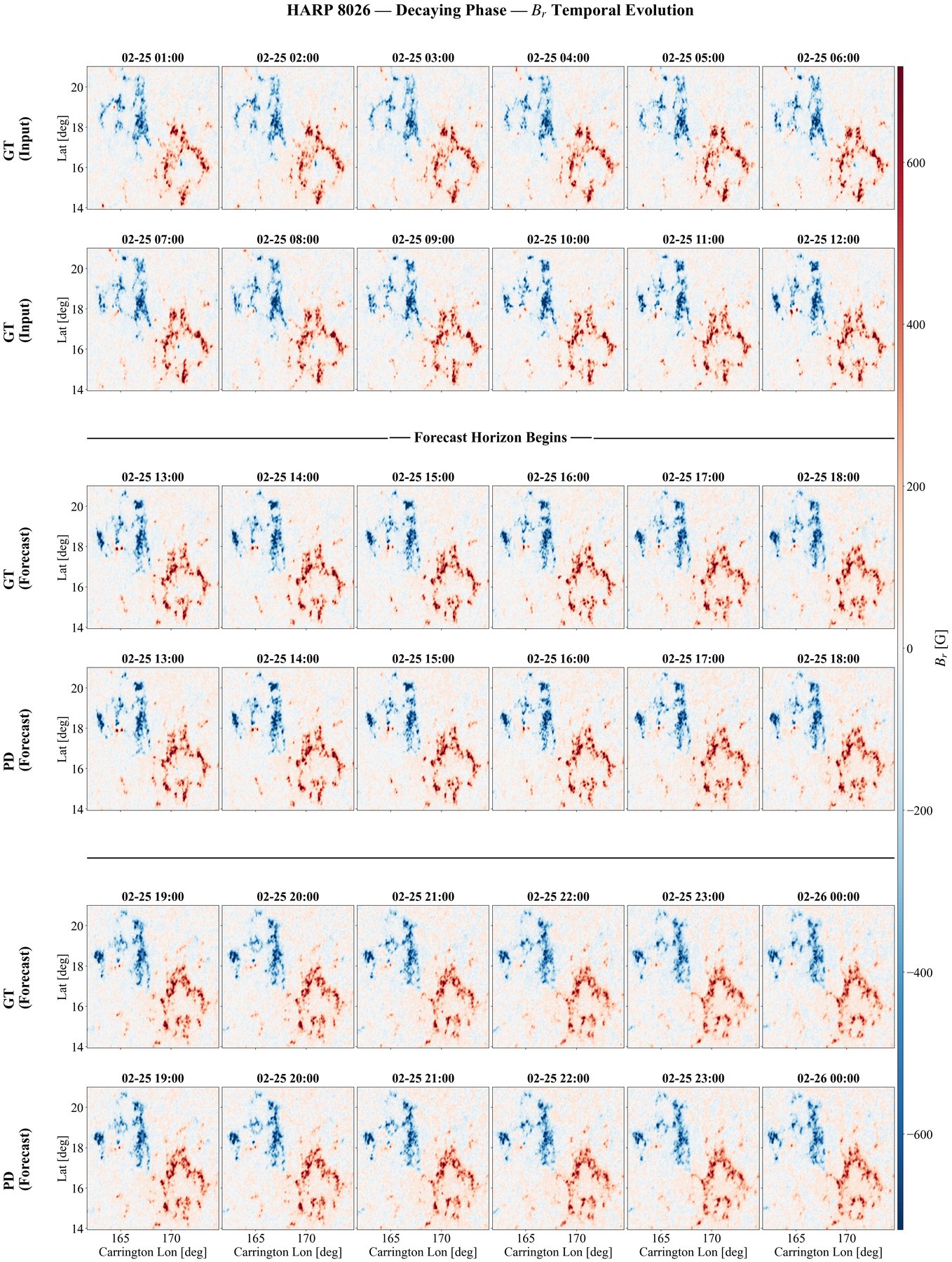}
\caption{Same as Figure~\ref{fig:app_8026_emerging} but for the \textbf{decaying} phase of HARP~8026.}
\label{fig:app_8026_decaying}
\end{figure*}

\clearpage

\section{Evolutionary Phase Comparison for \texorpdfstring{$B_p$}{Bp} and \texorpdfstring{$B_t$}{Bt} Components}\label{app:lifecycle_BpBt}

This appendix supplements the $B_r$ evolutionary phase comparisons shown in the main text (Figures~\ref{fig:ar7959_lifecycle} and~\ref{fig:ar8026_lifecycle}) by providing the corresponding results for the $B_p$ and $B_t$ components. The layout is identical: rows correspond to the emerging, steady, and decaying phases; columns show ground truth, predicted field, and residual at forecast hour~12. All images are displayed at original HMI pixel resolution with Carrington coordinate axes.

\begin{figure*}[p]
\centering
\includegraphics[width=0.95\textwidth]{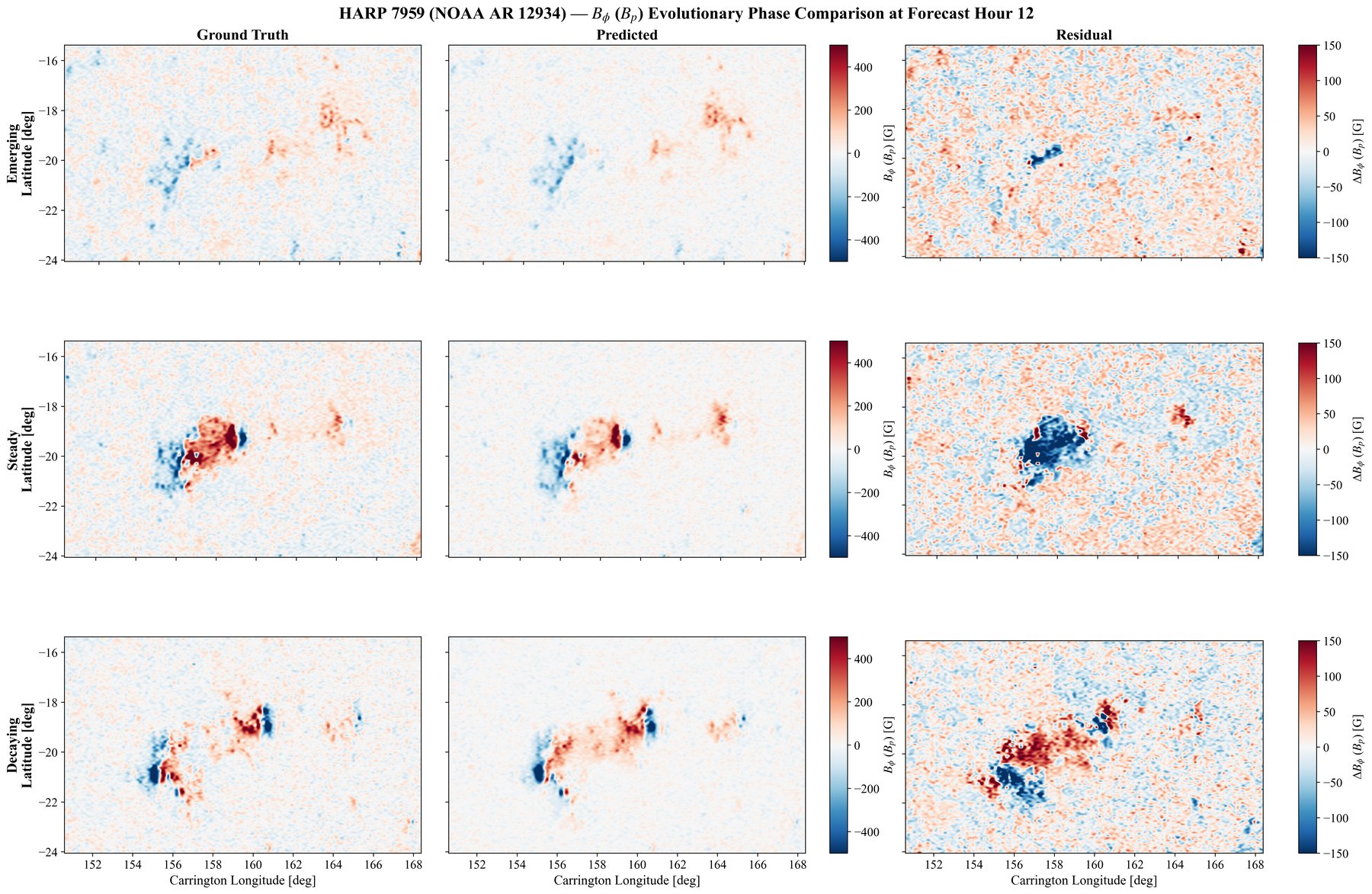}
\caption{$B_p$ evolutionary phase comparison for HARP~7959 at forecast hour~12. Layout is the same as Figure~\ref{fig:ar7959_lifecycle}.}
\label{fig:app_7959_lifecycle_Bp}
\end{figure*}

\begin{figure*}[p]
\centering
\includegraphics[width=0.95\textwidth]{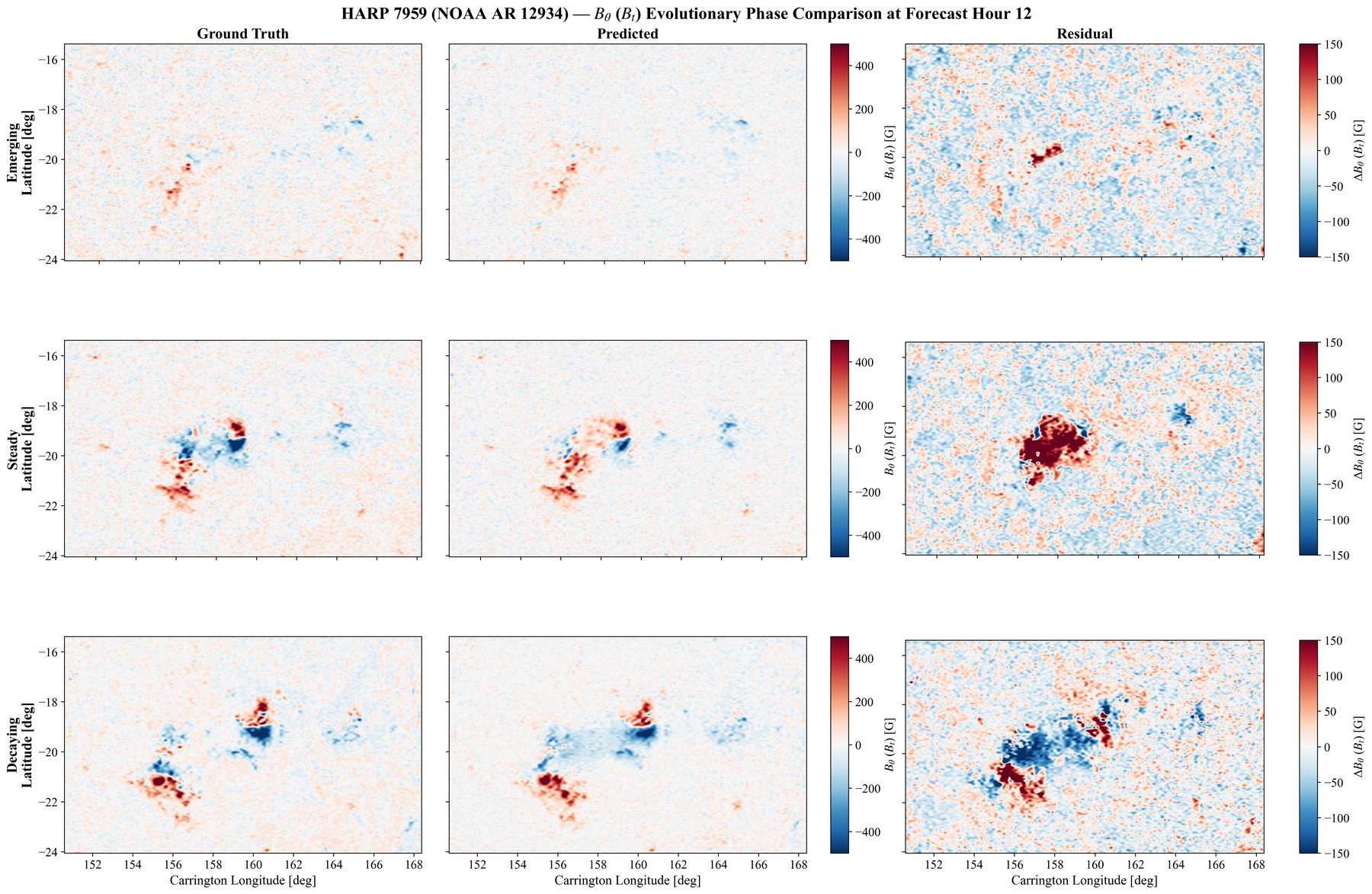}
\caption{$B_t$ evolutionary phase comparison for HARP~7959 at forecast hour~12. Layout is the same as Figure~\ref{fig:ar7959_lifecycle}.}
\label{fig:app_7959_lifecycle_Bt}
\end{figure*}

\begin{figure*}[p]
\centering
\includegraphics[width=0.95\textwidth]{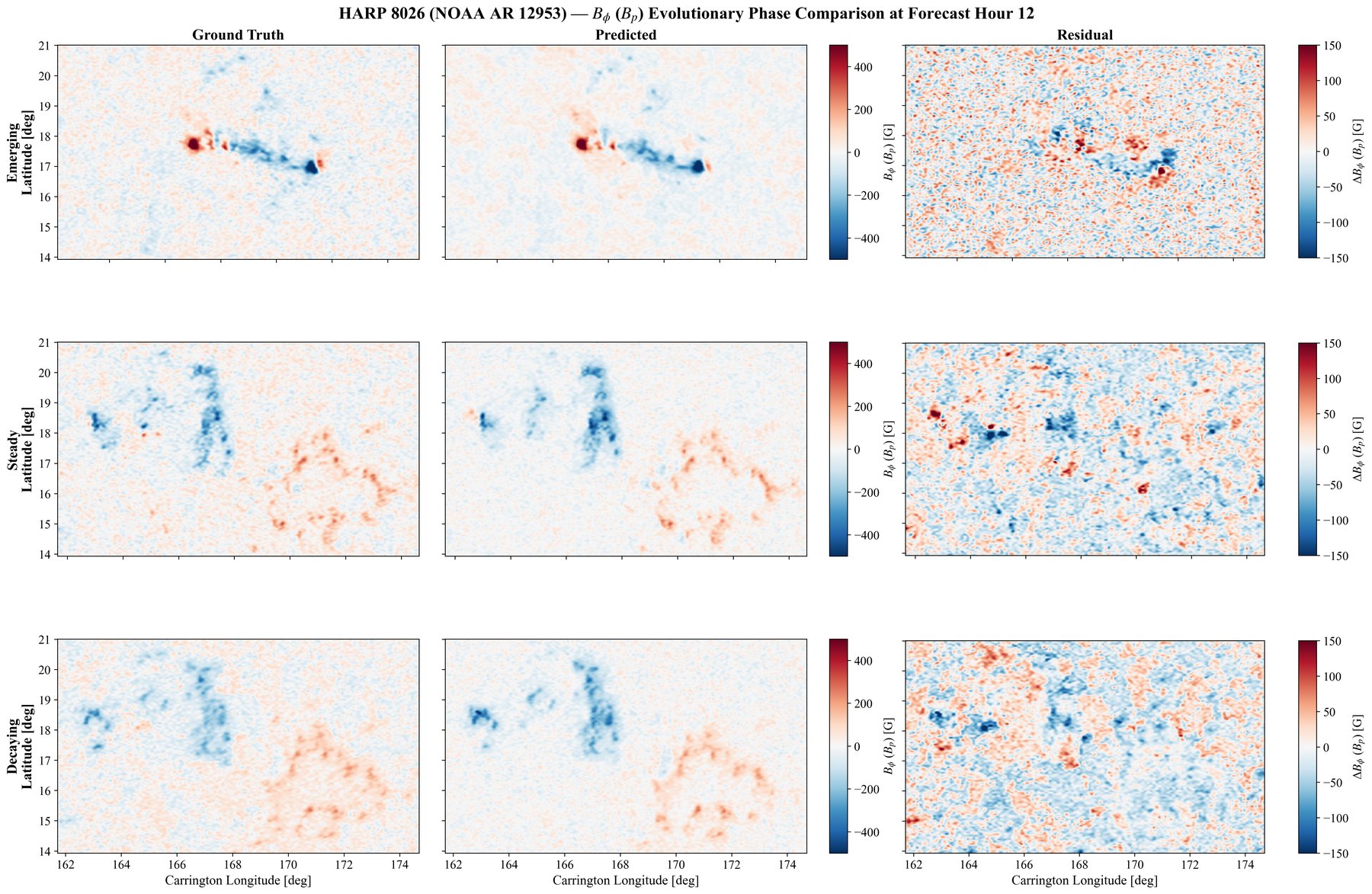}
\caption{$B_p$ evolutionary phase comparison for HARP~8026 at forecast hour~12. Layout is the same as Figure~\ref{fig:ar7959_lifecycle}.}
\label{fig:app_8026_lifecycle_Bp}
\end{figure*}

\begin{figure*}[p]
\centering
\includegraphics[width=0.95\textwidth]{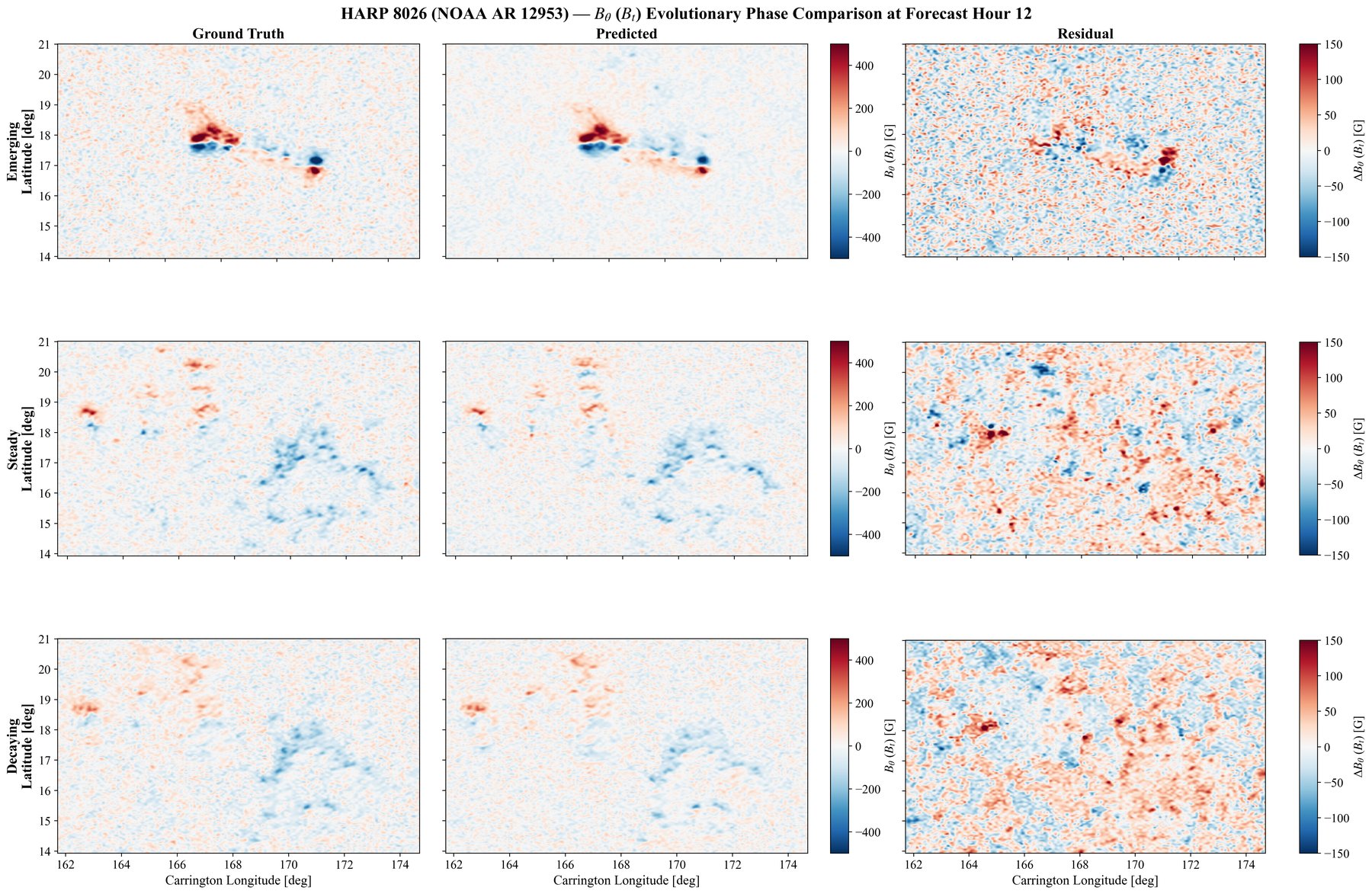}
\caption{$B_t$ evolutionary phase comparison for HARP~8026 at forecast hour~12. Layout is the same as Figure~\ref{fig:ar7959_lifecycle}.}
\label{fig:app_8026_lifecycle_Bt}
\end{figure*}

\clearpage

\bibliography{ref}
\bibliographystyle{plainnat}

\end{document}